\documentclass[journal]{vgtc}                

\ifpdf
  \pdfoutput=1\relax                   
  \usepackage{graphicx}                
  \DeclareGraphicsExtensions{.pdf,.png,.jpg,.jpeg} 
\else
  \usepackage{graphicx}                
  \DeclareGraphicsExtensions{.eps}     
\fi%

\graphicspath{{figures/}{pictures/}{images/}{./}} 

\usepackage{amsmath}
\usepackage{amssymb}
\usepackage{xifthen}
\usepackage{pgfplots}
\usepackage{tikz,graphicx}
\usetikzlibrary{arrows}
\usetikzlibrary{colorbrewer}
\usepgfplotslibrary{colorbrewer}

\tikzset{>=latex}

\newcommand{\todo}[2][]{%
  \ifthenelse{\equal{#1}{}}
    {{\color{red}TODO:~#2}}
    {{\color{red}TODO~[#1]:~#2}}
}

\makeatletter
\newcommand{\mypart@star}[2]{\part*{#1}\label{#2}}
\newcommand{\mypart@nostar}[3][]{\part[#1]{#2}\label{#3}}
\newcommand{\mychapter@star}[2]{\chapter*{#1}\label{#2}}
\newcommand{\mychapter@nostar}[3][]{\chapter[#1]{#2}\label{#3}}
\newcommand{\mysection@star}[2]{\section*{#1}\label{#2}}
\newcommand{\mysection@nostar}[3][]{\section[#1]{#2}\label{#3}}
\newcommand{\mysubsection@star}[2]{\subsection*{#1}\label{#2}}
\newcommand{\mysubsection@nostar}[3][]{\subsection[#1]{#2}\label{#3}}
\newcommand{\mysubsubsection@star}[2]{\subsubsection*{#1}\label{#2}}
\newcommand{\mysubsubsection@nostar}[3][]{\subsubsection[#1]{#2}\label{#3}}
\newcommand{\myparagraph@star}[2]{\paragraph*{#1}\label{#2}}
\newcommand{\myparagraph@nostar}[3][]{\paragraph[#1]{#2}\label{#3}}
\newcommand{\mysubparagraph@star}[2]{\subparagraph*{#1}\label{#2}}
\newcommand{\mysubparagraph@nostar}[3][]{\subparagraph[#1]{#2}\label{#3}}

\newcommand{\mypart}{\@ifstar{\mypart@star}{\mypart@nostar}}
\newcommand{\mychapter}{\@ifstar{\mychapter@star}{\mychapter@nostar}}
\newcommand{\mysection}{\@ifstar{\mysection@star}{\mysection@nostar}}
\newcommand{\mysubsection}{\@ifstar{\mysubsection@star}{\mysubsection@nostar}}
\newcommand{\mysubsubsection}{\@ifstar{\mysubsubsection@star}{\mysubsubsection@nostar}}
\newcommand{\myparagraph}{\@ifstar{\myparagraph@star}{\myparagraph@nostar}}
\newcommand{\mysubparagraph}{\@ifstar{\mysubparagraph@star}{\mysubparagraph@nostar}}
\makeatother

\usepackage{microtype}                 
\PassOptionsToPackage{warn}{textcomp}  
\usepackage{textcomp}                  
\usepackage{mathptmx}                  
\usepackage{times}                     
\usepackage{cite}                      
\usepackage{tabu}                      
\usepackage{booktabs}                  
\usepackage{url}
\usepackage{hyperref}
\usepackage{amsmath}
\usepackage{amssymb}
\usepackage{enumitem}
\usepackage{multirow}
\usepackage{color}

\onlineid{1180}

\vgtccategory{Technique}

\title{Void Space Surfaces\\to Convey Depth in Vessel Visualizations}

\author{Julian Kreiser, \textit{Student Member, IEEE}, Pedro Hermosilla, \textit{Member, IEEE}, and Timo Ropinski, \textit{Member, IEEE}}
\authorfooter{
\item Julian Kreiser~(Student Member, IEEE), Pedro Hermosilla~(Member, IEEE), and Timo Ropinski~(Member, IEEE) are with the Visual Computing Group, Ulm University. E-mail: julian.kreiser@uni-ulm.de and pedro-1.hermosilla-casajus@uni-ulm.de and timo.ropinski@uni-ulm.de.
}

\shortauthortitle{Kreiser \MakeLowercase{\textit{et al.}}: Void Space Surfaces}

\abstract{
To enhance depth perception and thus data comprehension, additional depth cues are often used in 3D visualizations of complex vascular structures. Accordingly, there is a variety of different approaches described in the literature, ranging from chromadepth color coding over depth of field to glyph-based encodings. Unfortunately, the majority of existing approaches suffers from the same problem. As these cues are directly applied to the geometry's surface, the display of additional information, such as other modalities or derived attributes, associated with a vessel is impaired. To overcome this limitation we propose \emph{Void Space Surfaces} which utilize the empty space in between vessel branches to communicate depth and their relative positioning. This allows us to enhance the depth perception of vascular structures without interfering with the spatial data and potentially superimposed parameter information. Within this paper we introduce \emph{Void Space Surfaces}, describe their technical realization, and show their application to various vessel trees. Moreover, we report the outcome of a user study which we have conducted in order to evaluate the perceptual impact of \emph{Void Space Surfaces} as compared to existing vessel visualization techniques.
} 

\keywords{Depth Perception, Void Space Surface, Chromadepth}

\CCScatlist{ 
 \CCScat{K.6.1}{Management of Computing and Information Systems}%
{Project and People Management}{Life Cycle};
 \CCScat{K.7.m}{The Computing Profession}{Miscellaneous}{Ethics}
}

\teaser{
  \centering
  \includegraphics[width=\linewidth]{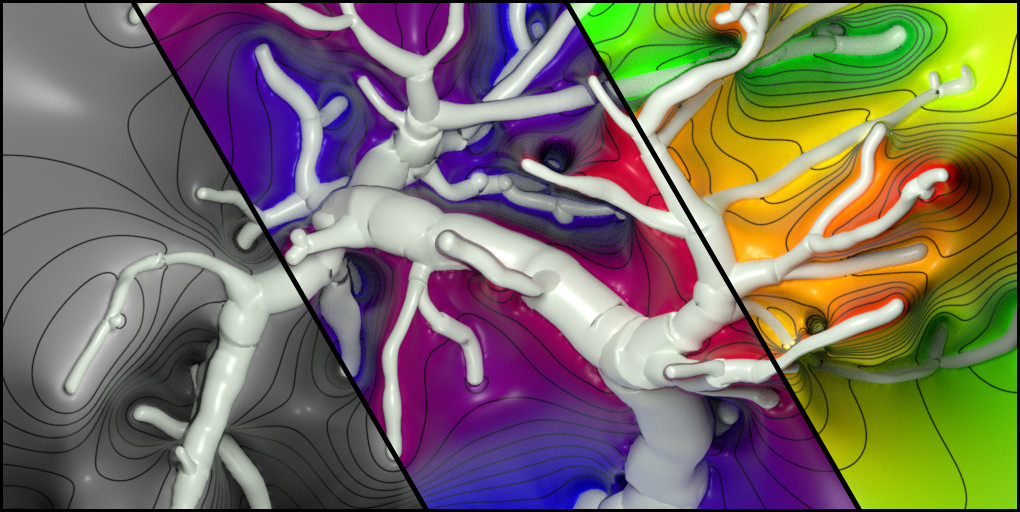}
  \caption{
	Example of a vascular structure visualized using our technique, Void Space Surfaces.
	This technique takes advantage of the empty space between vessels and interpolates a surface between them in which different depth cues can be applied, such as dark-means-depth (left), pseudo-chromadepth color scale (middle), chromadepth (right), or advanced illumination techniques (all figures).
	Moreover, since depth cues have been shifted away from the vessels' surface, additional information can be encoded on top of them.}
  \label{fig:teaser}
}

\renewcommand{\manuscriptnotetxt}{}

\vgtcinsertpkg
\usepackage{subcaption}

\begin{document}



\maketitle

\section{Introduction} 
Medical visualization is of key importance in many areas of medicine, such as surgery planning or education. Due to its relevance, researchers focused their attention on improving the perceptual capabilities of these visualizations~\cite{Preim2016}. Realistic lighting simulations~\cite{Magnus2018}, illustrative techniques~\cite{Bruckner2007} or different color schemes~\cite{Borkin2011} are just some of the approaches that have been proposed in order to improve shape or depth perception.

Regarding visual perception, visualization of vascular structures is one of the most challenging areas of medical visualization. Vascular structures are composed of numerous branches, which are complex by nature, and can easily clutter the resulting visualization. This makes it difficult to infer the relative position of the individual branches even when realistic lighting effects are used. To further complicate the situation, often additional information associated with the surface of a vessel needs to be communicated. Such information contains for instance additionally acquired or derived parameters such as blood flow, pressure or wall shear stress, which are typically conveyed by stream lines, different color scales or glyphs. Effectively communicating all this information without overwhelming the viewer is a problem that has been addressed by several authors in the past~\cite{Ropinski2006,Lawonn2014,Behrendt2017,Lichtenberg2017}.

Stereoscopic rendering is one of the most effective methods to communicate depth and shape of 3D objects. However, it usually requires instrumentation of the user, and is thus not widely available for most of the domain experts. Another commonly used technique to improve depth perception is chromadepth color coding~\cite{Richard1987,Bailey1998}. While with this technique, the perceptual benefits can be exploited when viewing appropriately coded images without special glasses, the effect can be further supported by using diffraction grating glasses. This fact was also exploited by the pseudo-chromadepth technique, which uses a color scale composed of only two colors (blue and red) to effectively convey depth information~\cite{Ropinski2006}. The authors could show, that the reduced number of colors significantly improved the depth perception as compared to chromadepth. While these approaches work well when communicating vessel structures, they do not allow for the communication of additional information, since they exploit the color channel to encode depth information. Thus, various approaches have been considered to overcome this problem. Behrendt~et~al.~\cite{Behrendt2017} introduced a technique to encode information on the vessels whilst maintaining the benefits of the pseudo-chromadepth color scale, by applying it on the edges of the vessels only. Lichtenberg et al.~\cite{Lichtenberg2017} used glyphs to communicate depth information on the vessel end-points freeing the surface of the vessel of such task. Despite these efforts to integrate the visualization of several vessel parameters into a single image, the problem cannot be considered as solved, as the applied depth perception are either exploiting the color channel also used to communicate the vessel structure or require the user to interpret glyphs in order to decipher depth relations. To our knowledge no monoscopic vessel visualization technique exists, which naturally encodes depth without occupying the color channel of the vessel's surface.

In this paper, we present \emph{Void Space Surfaces}~(VSS), a novel approach to visualize vascular structures. VSS are able to convey the depth of complex vessel structures without interfering with additional measures visually encoded on the vessels' surfaces. To do so, VSS take advantage of the empty space between vessels to generate camera dependent height fields, which serve as a canvas to convey depth (see Figure~\ref{fig:teaser}). Within this paper, we will show how to generate VSS, and how they can be used to convey the depth of vascular structures by applying the most commonly used depth cues, such as chromadepth, pseudo-chromadetph, occlusions, or dark-means-deep to these surfaces. To demonstrate the capabilities of VSS, we further have conducted a user study in which we compared VSS with previous vessel visualization techniques. Thus, our contributions are threefold:

\begin{itemize}[noitemsep,topsep=5pt,parsep=2pt,partopsep=0pt]
    \item First, we introduce \emph{Void Space Surfaces}~(VSS), a novel vessel visualization technique that improves the depth perception of complex vascular structures without interfering with the color channel used to communicate a vessel's geometry.
    \item Second, we show how VSS can be used as a canvas for well established depth cues, and thus leave room on the vessel's surface to communicate additional vessel parameters.
    \item Third, we report the results of a user study, which we have conducted in order to analyze the perceptual benefits of VSS as compared to previous vessel visualization techniques. 
\end{itemize}

The remainder of this paper is structured as follows. After having discussed related work in Section~\ref{sec:relatedwork} we describe general design goals for vessel visualizations, which have motivated our approach, in Section~\ref{sec:designgoals}. VSS are then introduced in Section~\ref{sec:vss} where we also show how they can be used to communicate depth. Section~\ref{sec:implementation} provides implementation details for a GPU-based real-time visualization of VSS. Section~\ref{sec:userstudy} discusses the results of our user study, whereas derived insights and limitations of VSS are discussed in Section~\ref{sec:discussion}. Finally, we will conclude in Section~\ref{sec:conclusions} and summarize our contributions and findings.

\section{Related Work}\label{sec:relatedwork}
In this section, we describe the work related to vessel visualization and depth perception enhancement in medical visualization. Since providing a comprehensive review of these topics is beyond our scope, we would like to refer the reader to the state-of-the-art report published by Preim et al.~\cite{Preim2016} which details many of these concepts.

\noindent\textbf{Medical visualization.} Several techniques have been proposed to improve the perception of medical visualizations. Global illumination techniques, for example, have been proven to improve the perception of surface features in volume visualization~\cite{Zheng2014}. Ropinski et al.~\cite{Ropinski2008} and Diaz et al.~\cite{Diaz2010b} for instance developed different techniques to approximate ambient occlusion in volumetric datasets in order to improve shape and depth perception. In a more general setup, Wanger et al.~\cite{Wanger1992} demonstrate how shadows can significantly improve the perception of the relative position of objects in a scene. Solteszova et al.~\cite{Solteszova2010} developed an illumination model based on the work of Schott et al.~\cite{Schott2009} which is able to generate realistic shadows in volume rendering. Besides simulating realistic lighting effects, illustrative techniques have also been used to improve perception in medical visualization. Bruckner et al.~\cite{Bruckner2007} proposed Style Transfer Functions, which realize non-photorealistic rendering in the context of volume visualization. Šoltészová et al.~\cite{Solteszova2011} apply chromadepth to improve shadow perception in volume visualization, and Ebert et al.~\cite{Ebert2000} applied different illustrative techniques such as halos, boundary enhancement, or silhouettes in volume visualization.

Maximum Intensity Projection (MIP) is a rendering technique often used to visualize contrast-enhanced data. However, with this technique, due to the lack of shading, is difficult to perceive the depth of the individual objects and their relative positions. Diaz et al.~\cite{Diaz2010} presented a technique to improve depth perception on MIP images. Therefore, they compute a depth value while traversing the volume dataset, and use this value to modify the final color of the MIP image. Bruckner et al.~\cite{Bruckner2009} incorporated shading into MIP, improving depth perception and allowing a smooth transition between MIP and direct volume rendering.

\noindent\textbf{Vessel visualization.} Vessel datasets are often represented as a volume and then visualized using direct volume rendering~\cite{Kubisch2012}. Therefore, vessel visualization can take advantage of existing volume visualization techniques to improve shape and depth perception. However, representing such data as a volume is not the only approach used by domain experts, as volume datasets often include other tissues that have to be filtered before visualizing the vascular structures. Thus, several algorithms exist for extracting the vascular structures from volume datasets and render them by using different primitives (such as meshes or truncated cones)~\cite{Gerig1993, Hahn2001, Oeltze2005}. Due to the different representations used and the particular shape of these structures, specific techniques were developed to improve perception in such models. Ritter et al~\cite{Ritter2006} developed an illustrative technique which uses hatch lines~\cite{Praun2001} on top of the vessels to encode occlusions between the different branches of the structure. Moreover, they also applied a stroke texture with a varying width depending on the distance to the viewer. Lawonn et al.~\cite{Lawonn2015} presented a technique which generates illustrative shadows with support lines and silhouettes to improve perception on vascular structures. Moreover, they also proposed a 2D visualization in which distance was encoded by different hatching styles. Lichtenberg et al.~\cite{Lichtenberg2017} presented a different approach to communicate depth in vessel visualizations. They drew view-dependent circle glyphs on vessel end-points which encoded depth information. Moreover, they also proposed an algorithm to detect these points in vascular structures.

Whilst all these methods are focused on rendering extra information on top of the vessels, other techniques have centered their attention on communicating depth information through coloring the vessels.
Ropinski et al.~\cite{Ropinski2006} used a color scale based on the chromadepth technique composed of only two colors which encodes depth information. This color scale, named pseudo-chromadepth, performed better than other techniques in the user study carried out by the authors, as it resulted in more accurate and faster responses. Joshi et al.~\cite{Joshi2008} presented a set of techniques to improve perception in visualizations of vascular structures represented by a volume data set. The authors used a similar color encoding as the one used in pseudo-chromadepth together with other illustrative visualization techniques such as tone shading or halos. In the same year, Chu et al.~\cite{Chu2008} presented a technique which combined a similar color encoding to the one used in the chromadepth technique with hatching rendering and silhouettes. Kersten-Oertel et al.~\cite{Kersten2014}, in 2014, carried out a user study in which they compared the effect of different depth cues on vessel visualizations. Based on the results of their study, they conclude that pseudo-chromadepth and the use of fog achieve the best improvement in depth perception.

Encoding depth information as colors on the vessel surfaces has been proven to be effective in communicating depth. However, these techniques limit the information presented on the vessels to depth information. Domain experts are often interested in visualizing also other properties on top of the vessels.
This problem was addressed by Behrendt et al.~\cite{Behrendt2017} who combined pseudo-chromadepth with additional information on top of the vessels. They used the pseudo-chromadepth color scheme to shade the areas close to the contour by applying a blending mask inspired by Fresnel equations. To represent the additional information they used a discretized color scale which reduced ambiguity between the surface color and the pseudo-chromadepth scale. Lawonn et al.~\cite{Lawonn2014} combined different techniques to convey depth together with vessel flow visualization, accurately resolving at the same time occlusions between different vessels. Borkin et al.~\cite{Borkin2011} used the surface of the vessel together with a 2D representation of them to communicate endothelial shear stress.

\section{Design Goals for Vessel Visualization}\label{sec:designgoals}
The visualization techniques proposed in this paper have been developed with certain design goals in mind. While these goals are formulated with the application of vessel visualization in mind, we believe that many of them are also relevant for medical visualization in general, and beyond.

\noindent\textbf{Morphological features.} When inspecting complex 3D structures such as vessel trees, it is of uttermost importance that depth and shape are communicated effectively. Especially for vessels, these two parameters are intervened. Adequately communicating the depth of various vessel branches also facilitates the communication of the overall shape of the vascular tree. Besides these global shape parameters, also the details on the surface of a vessel need to be communicated, as these might give hints to initial aneurysms or other abnormalities. Consequently, depth and shape of a vessel need to be communicated efficiently to be able to decipher the structure of a vessel.

\noindent\textbf{Functional parameters.} Apart from these morphological features, often functional features are relevant in vessel visualization. Such parameters include wall shear stress and pressure in hemodynamics, elasticity, occurrence of plaque, or vascular constrictions, just to name a few. As in many subareas of medical visualization, where multiparametric or multimodal data needs to be visualized, it is essential to visualize these functional parameters in the context of the morphology, as only then the association between function and structure becomes possible. Consequently, most existing visualization techniques convey the functional parameters as being overlaid over the vessel. While this naturally interferes with the communication of surface details through shading, research related to color constancy suggests, that this proceedings still maintains a reasonable perception of the individual objects~\cite{foster2011color}. Nevertheless, the mapping of functional parameters exploits the same color channel as already used to communicate the vessel's surface structure. Therefore, it can not efficiently be exploited to communicate additional depth cues.

\noindent\textbf{Instrumentation avoidance.} Vessel visualization should ideally not rely on an instrumentation of the viewer, i.e., they should be effective without wearing glasses or using other aids. While stereoscopic rendering techniques could be considered as being very effective in communicating depth, they require an instrumentation of the user via stereo glasses and are thus less desirable for medical visualizations. Furthermore, along a similar line, ideally a vessel visualization technique also works on static images. This is important, since often images are parts of medical reports, which are still today often printed and exchanged on paper.

\noindent\textbf{Intuitive communication.} Finally, it is important that vessel visualizations can be intuitively interpreted. While it is certainly possible to craft dedicated visualizations, which precisely communicate the depth of certain structures, it is not clear that such techniques are intuitively interpretable. Furthermore, it can be expected that such techniques add to the visual clutter, and might thus even result in cognitive overload. Therefore, we assume that vessel visualizations should be natural in a way that they exploit natural depth cues, which do not require additional explanations to be interpreted.

\section{Void Space Surfaces for Vessel Visualization}\label{sec:vss}
Following our guidelines discussed above, and motivated by the usually large amounts of empty space between vessel structures, we utilize this area for the enhancement of depth and shape perception. To do so, we propose void space surfaces~(VSS), which allow us to completely eliminate all depth cues from the vessel surface's color channel and shift them to regions which are usually unused. As a consequence, we take advantage of the entire screen by using the void space for depth enhancements. Thus, VSS have been designed to improve the communication of morphological structures, while at the same time allowing for mapping functional parameters onto the vascular surface without interference.

We do not differentiate between mesh or volume representations for vessel visualizations. Our only requirement to generate a VSS is the availability of a depth map which exhibits enough void space to be effectively utilized. While, based on our design goals, we tailored our visualization technique towards both, static 2D images and 3D interactive applications, it has to be considered that for example for 2D printouts not all depth cues can be applied to VSS. Such an example would be for instance the parallax effect.

The following subsections describe how we generate VSS and the visualization techniques we use to enhance depth and shape perception.


\subsection{Surface Generation}
VSS are essentially view dependent height fields synthesized between vascular structures. To generate such a field, we interpolate the depth values of each pixel along the object's silhouette which serves as a basis for subsequent visualization techniques. During interpolation, we also limit the amount of contour points considered for each pixel to the silhouette contours the pixel is enclosed by, in order to avoid a contribution of unrelated structures.

The way the interpolated depth values are inferred for VSS is in principle arbitrary since there is no true solution to this problem, as VSS are an artificial construct. However, the VSS should be smooth to avoid the introduction of visual artifacts.

In our case we use \emph{Inverse Distance Weighting}~\cite{shepard1968two} which is smooth and has a control parameter to adjust the influence of samples depending on their distance to the target point~(see Equations~\ref{eq:isd}). Other spatial interpolation techniques such as \emph{Green Coordinates}~\cite{lipman2008green}, \emph{Poisson Image Editing}~\cite{perez2003poisson}, or \emph{Kriging Interpolation}~\cite{oliver1990kriging} might also be useable to fill the empty areas between vascular structures. However, \emph{Inverse Distance Weighting} has the advantages that it is easy to implement and each pixel can be calculated in parallel which makes it perfectly suitable to achieve interactive frame rates through a GPU implementation. Thus, we synthesize the depth values as follows:

\begin{align}
    \begin{split}
	    z'(x_0) &= \frac{\sum_{i=1}^{N} w(x_i) \cdot z(x_i)}{\sum_{i=1}^{N} w(x_i)} \\
    	w(x_i) &= \frac{1}{d(x_0, x_i)^p}
    \end{split}
    \label{eq:isd}
\end{align}

\noindent whereby $z'$ is the depth at an interpolated point $x_0$ and $z$ denote the sample depth at an interpolating point $x_i$ with $N$ being the total number of points. The weight $w$ consists of the inverse of a given distance metric $d$ from $x_0$ to $x_i$. The so called power parameter $p$ controls the amount of applied smoothing.

\begin{figure}[!t]
  \begin{subfigure}[t]{0.49\linewidth}
      \centering
      \frame{\includegraphics[width=\linewidth]{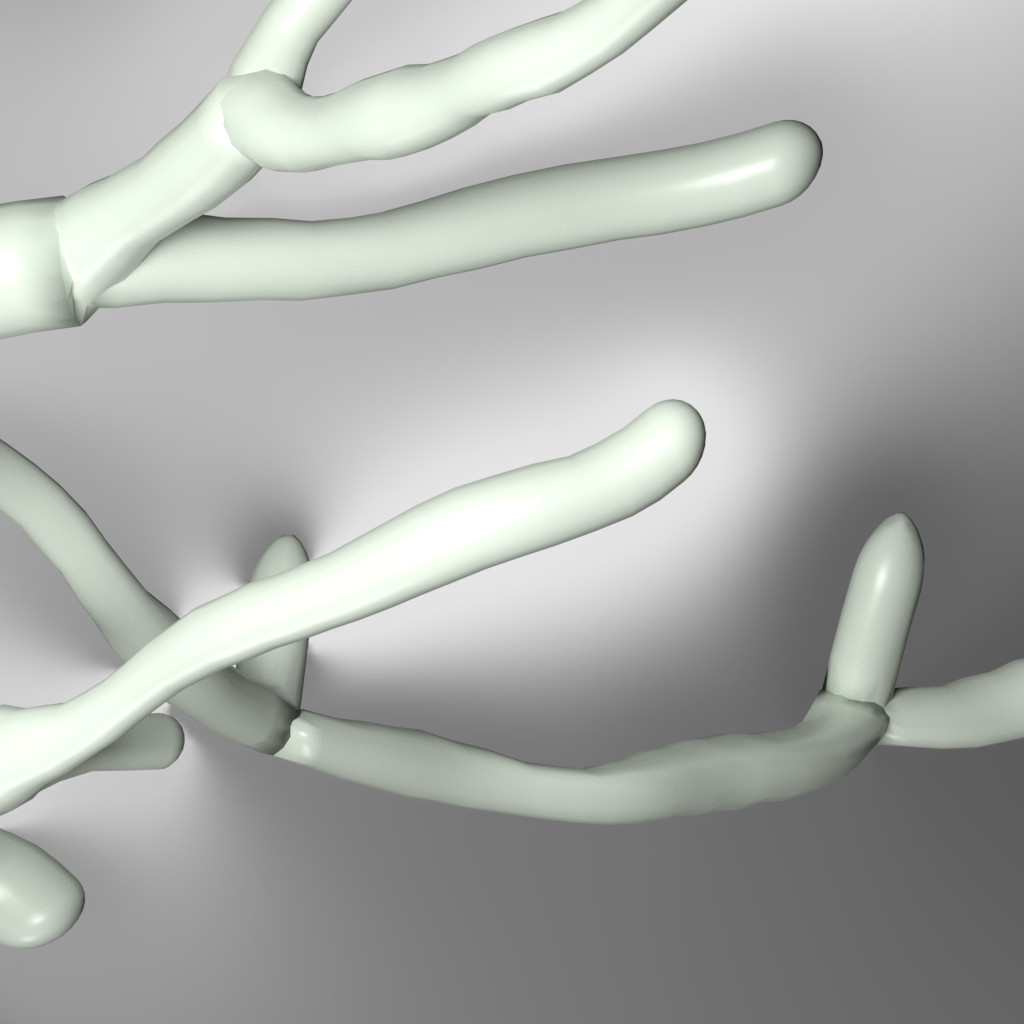}}
      \caption{Vessels without AO applied.}
  \end{subfigure}
  \hfill
  \begin{subfigure}[t]{0.49\linewidth}
      \centering
      \frame{\includegraphics[width=\linewidth]{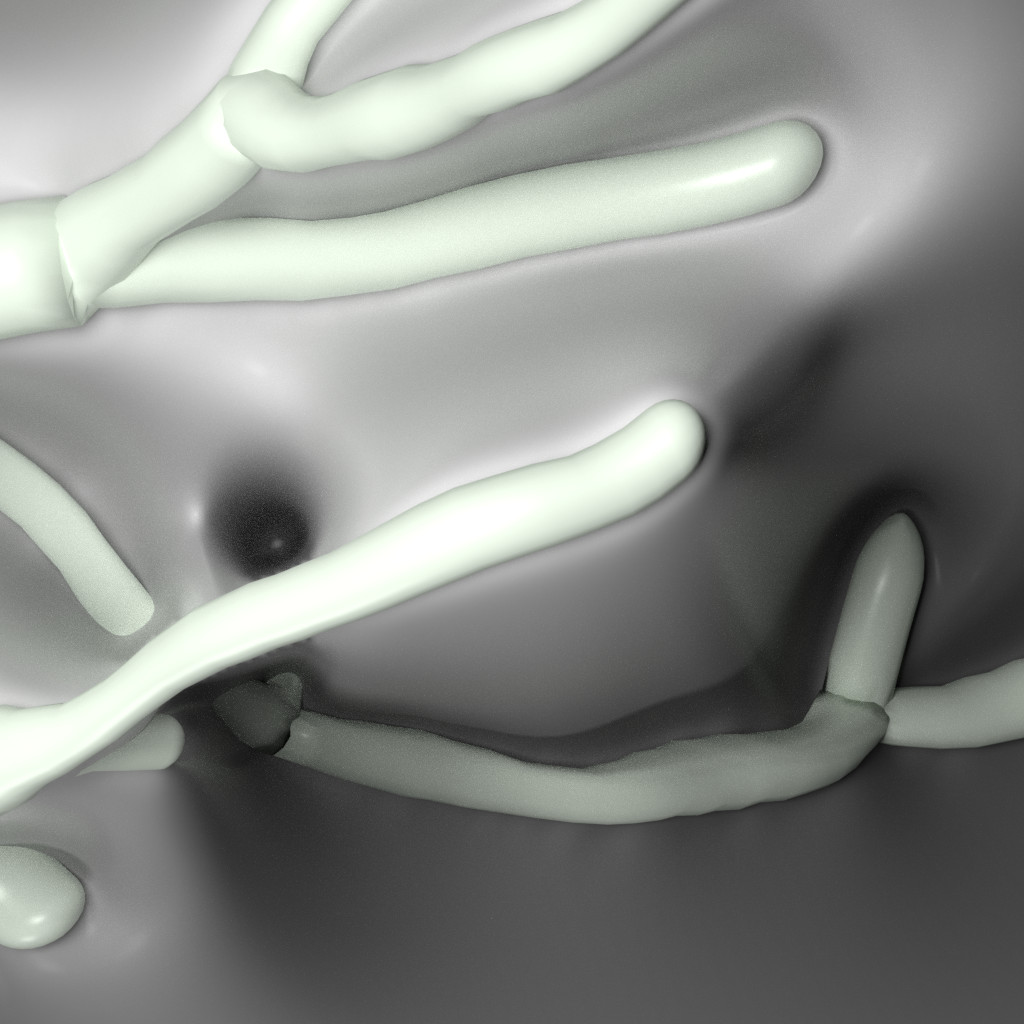}}
      \caption{Vessels with AO applied.}
  \end{subfigure}
  \caption{
	These figures illustrate the same VSS without~(a) and with~(b) depth anchoring. 
	The anchoring effect, which is achieved by applying ambient occlusion, intuitively \emph{glues} the vessel structure onto the VSS and thus enables an intuitive transfer of depth information.}
  \label{fig:anchoring}
\end{figure}

\noindent\textbf{Depth anchoring.} To enable VSS to become a canvas for communicating depth of the visualized vessel trees, it is crucial that the vessel tree is visually associated with the VSS. Only through such an association, to which we refer as depth anchoring, the depth as communicated through the VSS is intuitively transferred to the vessel structure. To realize depth anchoring in the context of our design guidelines, we have decided to exploit natural illumination effects. Furthermore, to be able to perform depth anchoring at interactive frame rates, we use a screen space ambient occlusion technique~\cite{ritschel2009approximating}. Figure~\ref{fig:anchoring} shows the effect of depth anchoring by visualizing a VSS without~(a) and with depth anchoring~(b) achieved through ambient occlusion. It can be clearly seen that the vessel structures seem to be aligned along the VSS when enabling depth anchoring. This in turn results in an intuitive transfer of depth information between the VSS and the vessel structure.

\begin{figure}[!t]
  \begin{subfigure}[t]{0.49\linewidth}
      \centering
  		\frame{\includegraphics[width=\linewidth]{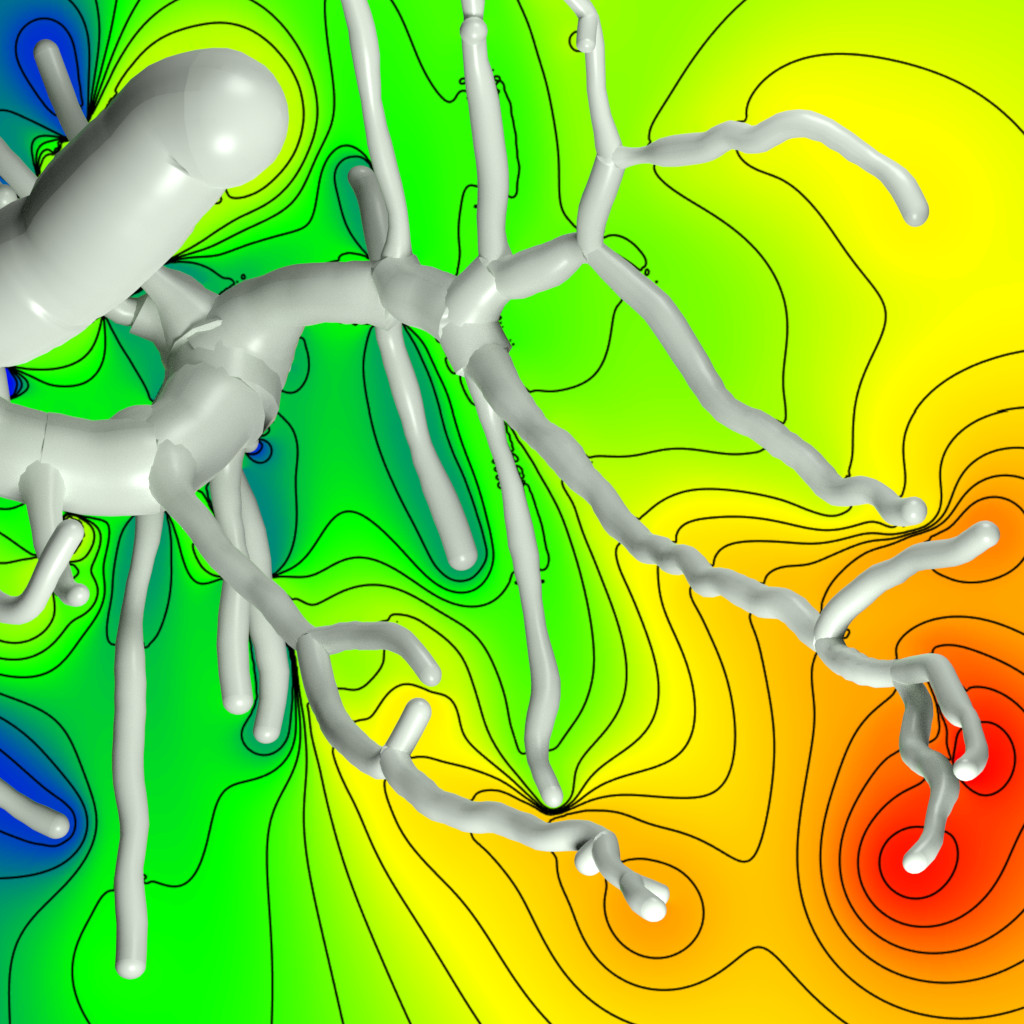}}
  \end{subfigure}
  \begin{subfigure}[t]{0.49\linewidth}
      \centering
  		\frame{\includegraphics[width=\linewidth]{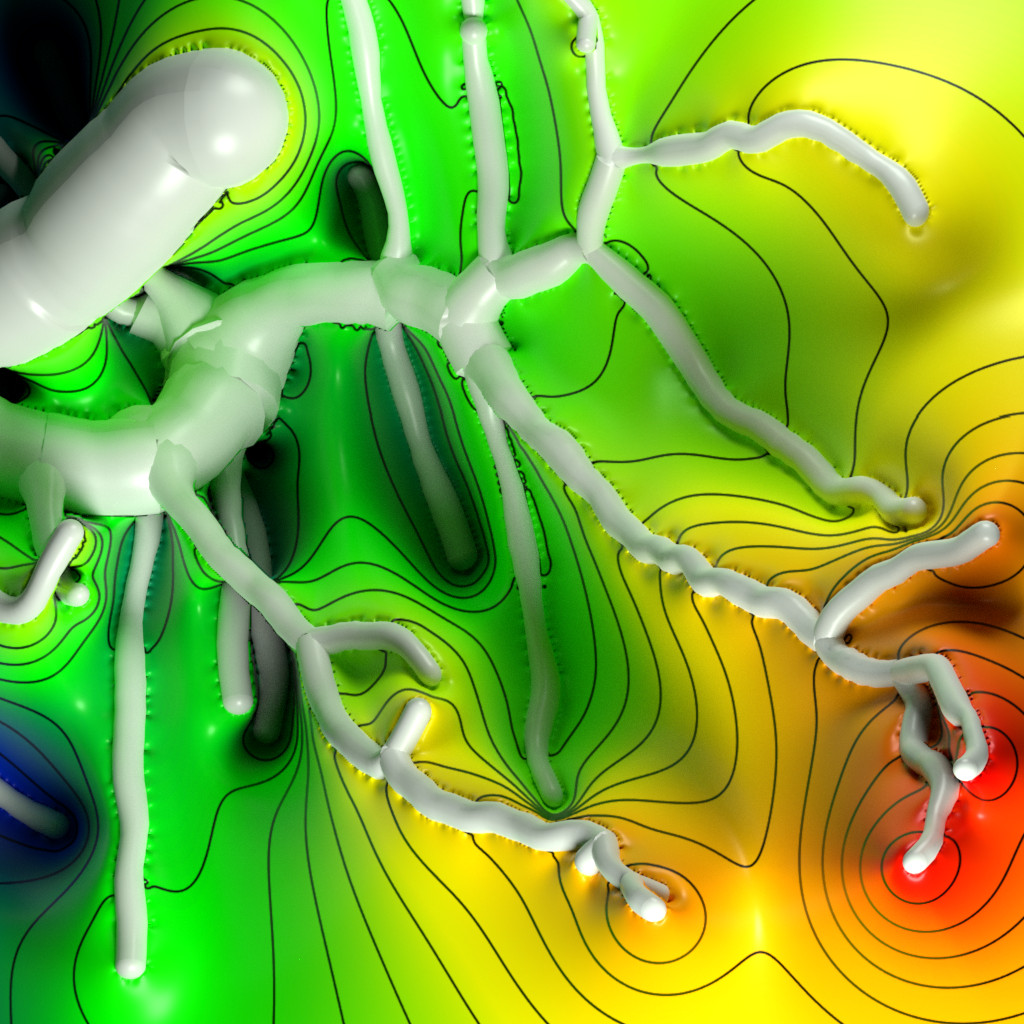}}
  \end{subfigure}
  
  \vspace{0.1\baselineskip}
  
  \begin{subfigure}[t]{0.49\linewidth}
      \centering
  		\frame{\includegraphics[width=\linewidth]{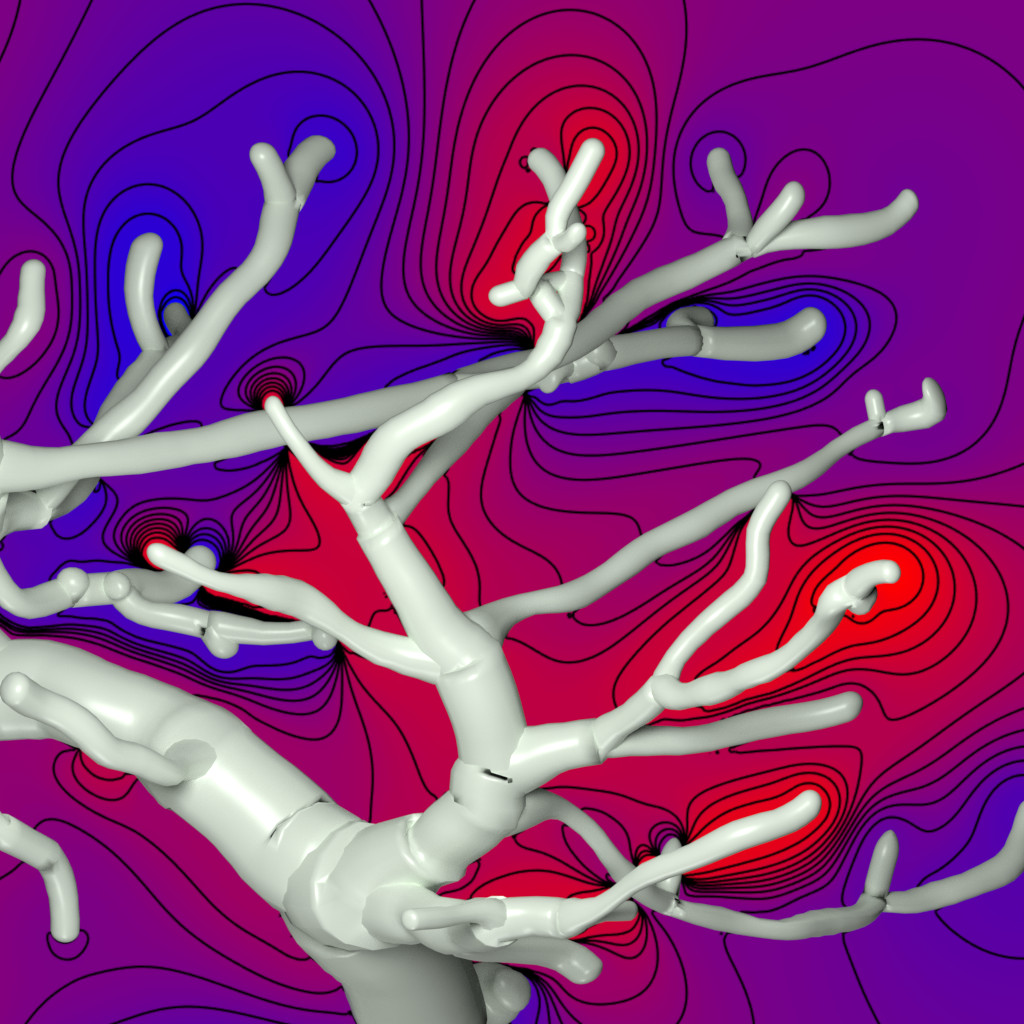}}
  \end{subfigure}
  \begin{subfigure}[t]{0.49\linewidth}
      \centering
  	\frame{\includegraphics[width=\linewidth]{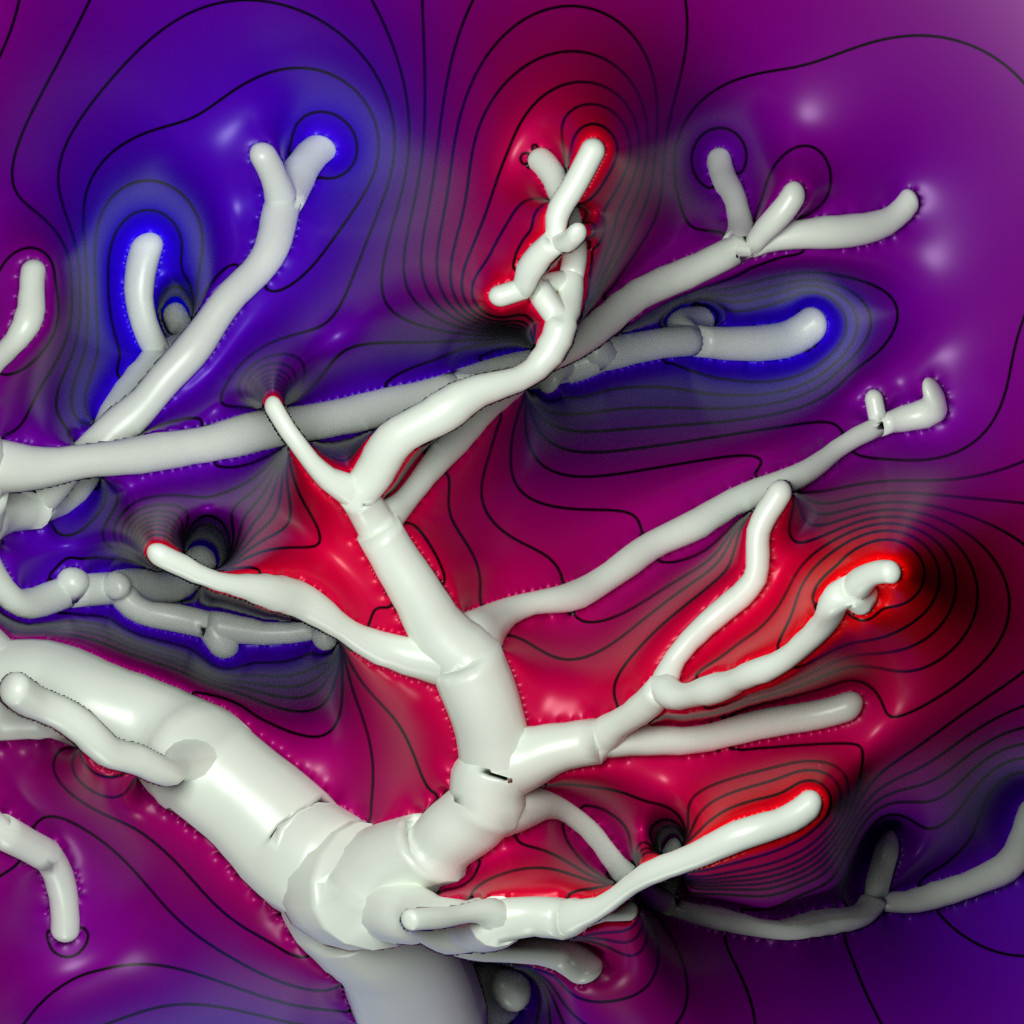}}
  \end{subfigure}
  
  \caption{Even with depth cues such as pseudo-chromadepth or iso-lines, illumination is one of the most effective ways to convey depth and shape. In these figures, we illustrate the improvement in depth perception achieved by shading the VSS using a global illumination algorithm~(right) over flat shaded background depth cues~(left).}
  \label{fig:illumination}
\end{figure}

To establishes a clear visual connection between the VSS and the vascular surface model, shading of the VSS plays also a major role. If the vessel geometry and the VSS are illuminated with the same lighting settings, the depth impression and scene comprehension is greatly improved. Reflections on the VSS accentuate cavities, hills, and neighboring occlusion which support the perception of relative and global depth differences. Either calculated with global illumination algorithms or simplified shading models, the perceived surface is immediately more vivid as demonstrated in Figure~\ref{fig:illumination}. Without any highlights, the VSS appears rather as a flat wallpaper-like background image which requires cognitive interpretation efforts to decipher depth.

\begin{figure}[!t]
  \begin{subfigure}[t]{0.49\linewidth}
      \centering
      \frame{\includegraphics[width=\linewidth]{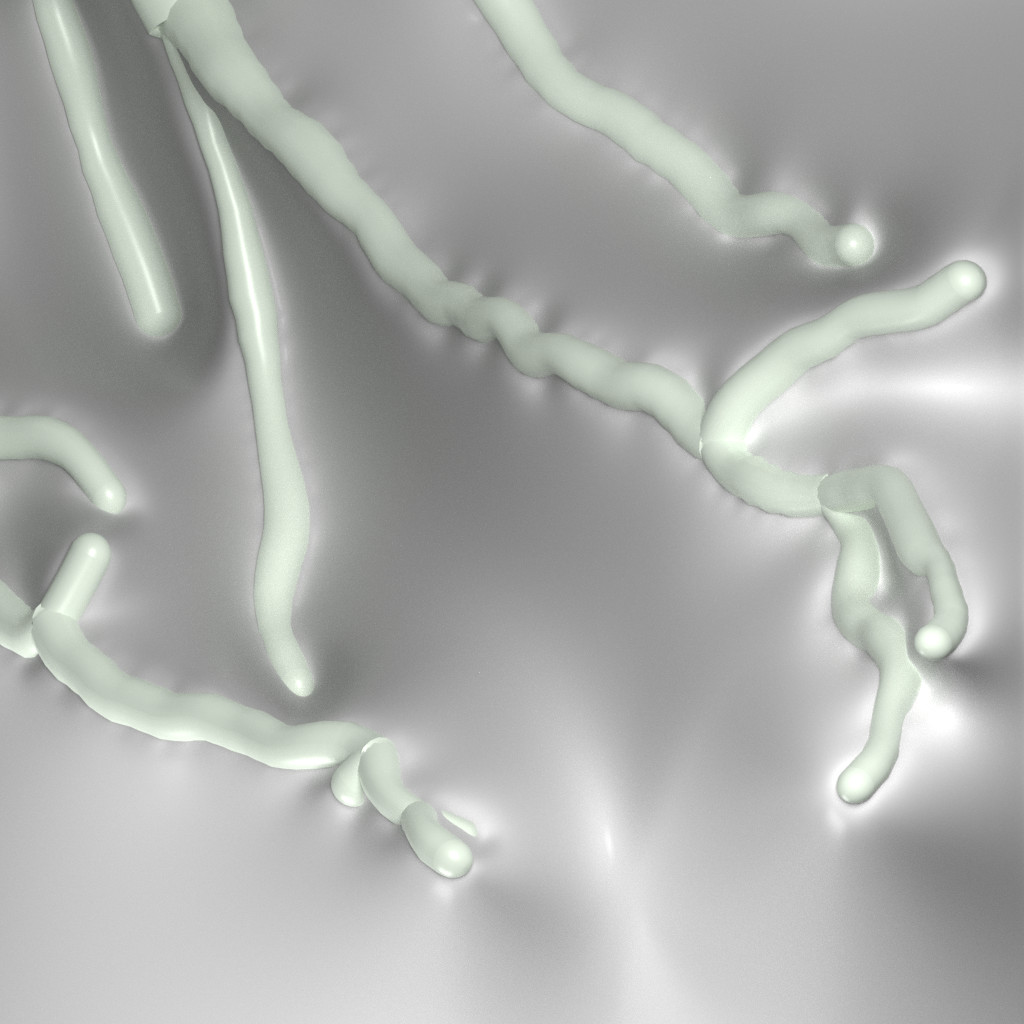}}
  \end{subfigure}
  \hfill
  \begin{subfigure}[t]{0.49\linewidth}
      \centering
      \frame{\includegraphics[width=\linewidth]{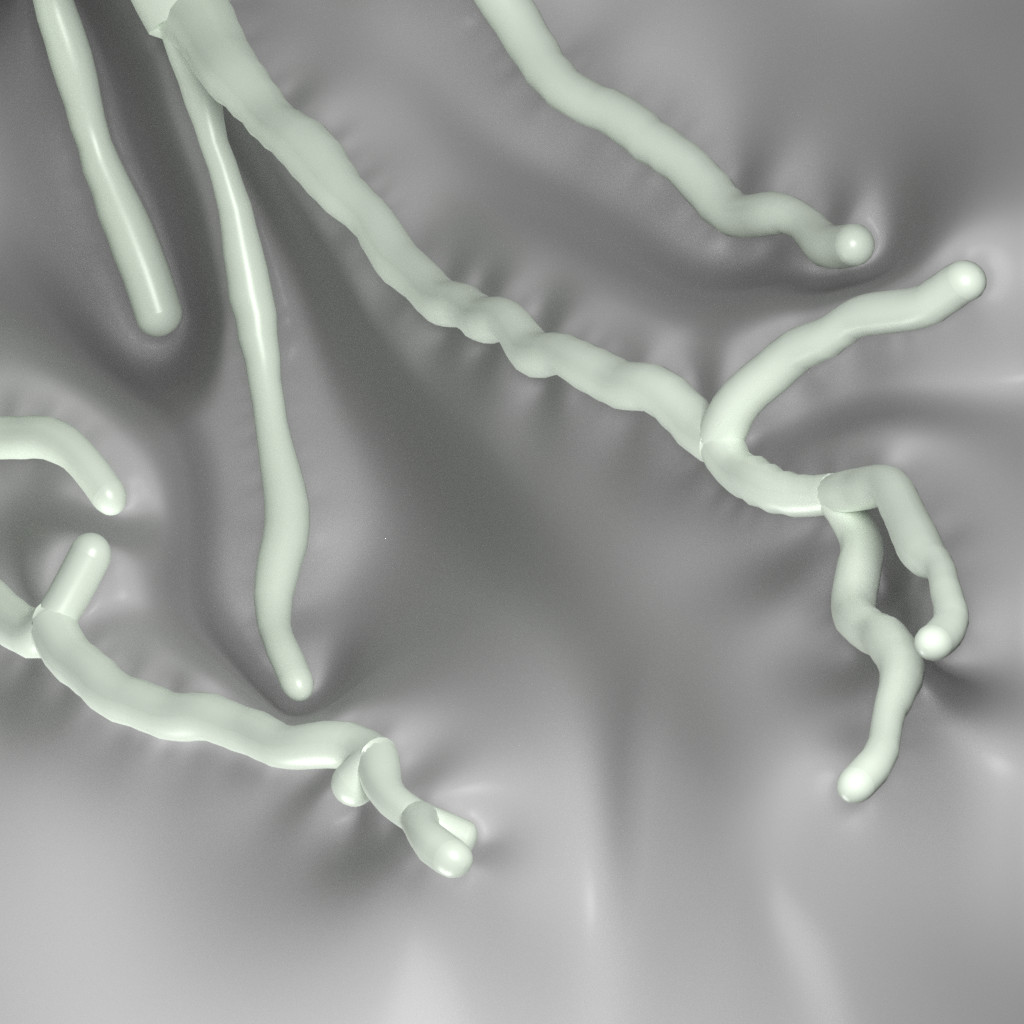}}
  \end{subfigure}
  \caption{
	The effect of parameter $p$ on the interpolated depth values is illustrated in these figures.
	High values of $p$ (right) increase the weight of local features while distant depth values almost do not contribute to the interpolated depth, generating sharp transitions.
	Lower values, on the contrary, create smooth transitions between the vessels (left).
	}
  \label{fig:power_parameter}
\end{figure}

\noindent\textbf{Power parameter variation.} When constructing VSS, greater values of $p$ result in a greater influence of samples close to the interpolated point, see Figure~\ref{fig:power_parameter}. For areas further away from the contour and without a distinct closest structure, the result is often a mostly averaged interpolated depth across the whole boundary. This is especially the case if a lot of contour points contribute to a pixel or the depth values along the contour are from a wide range. Local features can be preserved further into open areas with an increasing power parameter $p$ for the interpolation which sharpens the boundaries between different levels of depth. Thus, in the most extreme case, the VSS develops patterns similar to a Voronoi diagram.

If the vascular model does not cover a lot of screen space area, most of the surrounding pixels closer to the image border approach the average value of the outermost silhouette. This forms a fairly dull surface which does not communicate a lot of depth and shape of the vascular model. Vessels which intersect the image border separate the VSS into multiple regions which describe their surrounding contours much more accurate. Generally speaking, for a meaningful overview and a significant global depth and shape impression, the vascular model should be kept as close as possible to the image boundary.

\subsection{Perceptual Enhancement}
\begin{figure*}[!t]
  \begin{subfigure}[t]{0.24\linewidth}
      \centering
      \frame{\includegraphics[width=\linewidth]{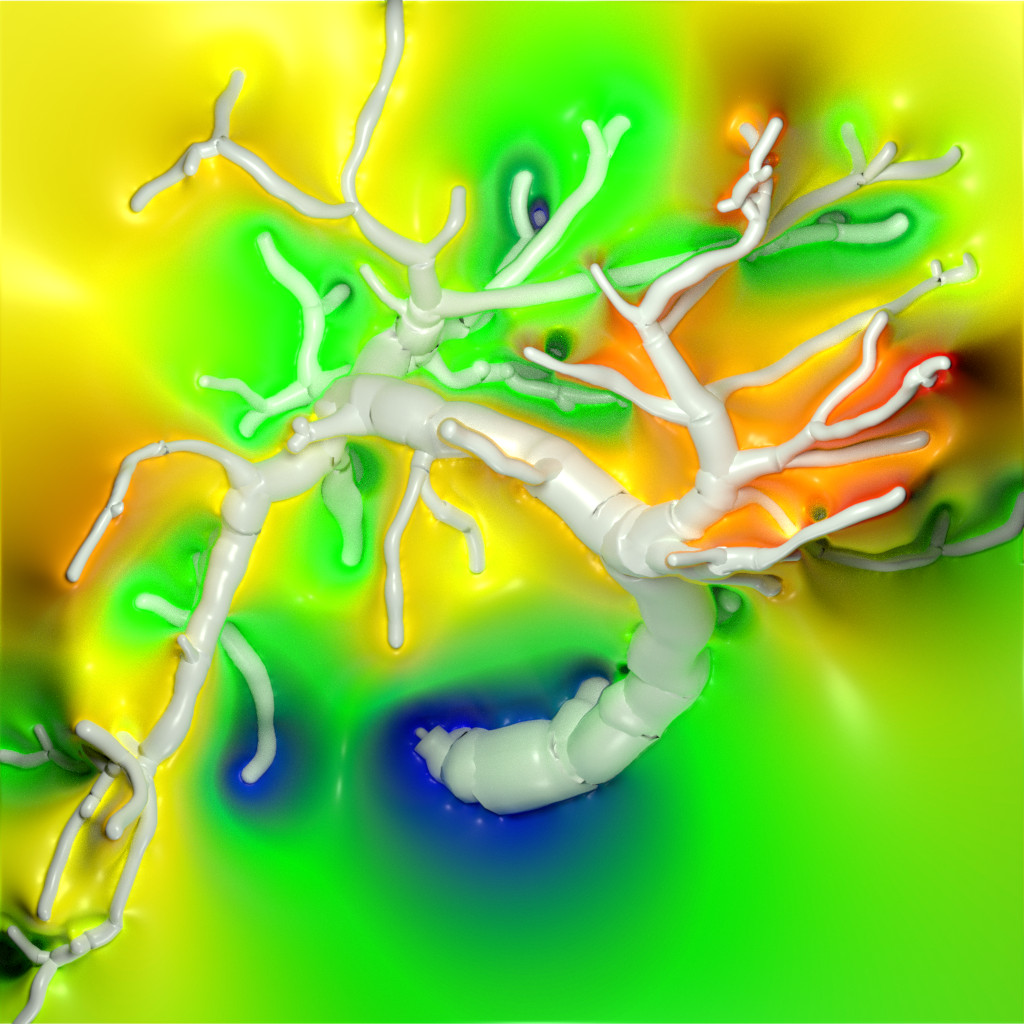}}
			\caption{}
  \end{subfigure}
  \hfill
  \begin{subfigure}[t]{0.24\linewidth}
      \centering
      \frame{\includegraphics[width=\linewidth]{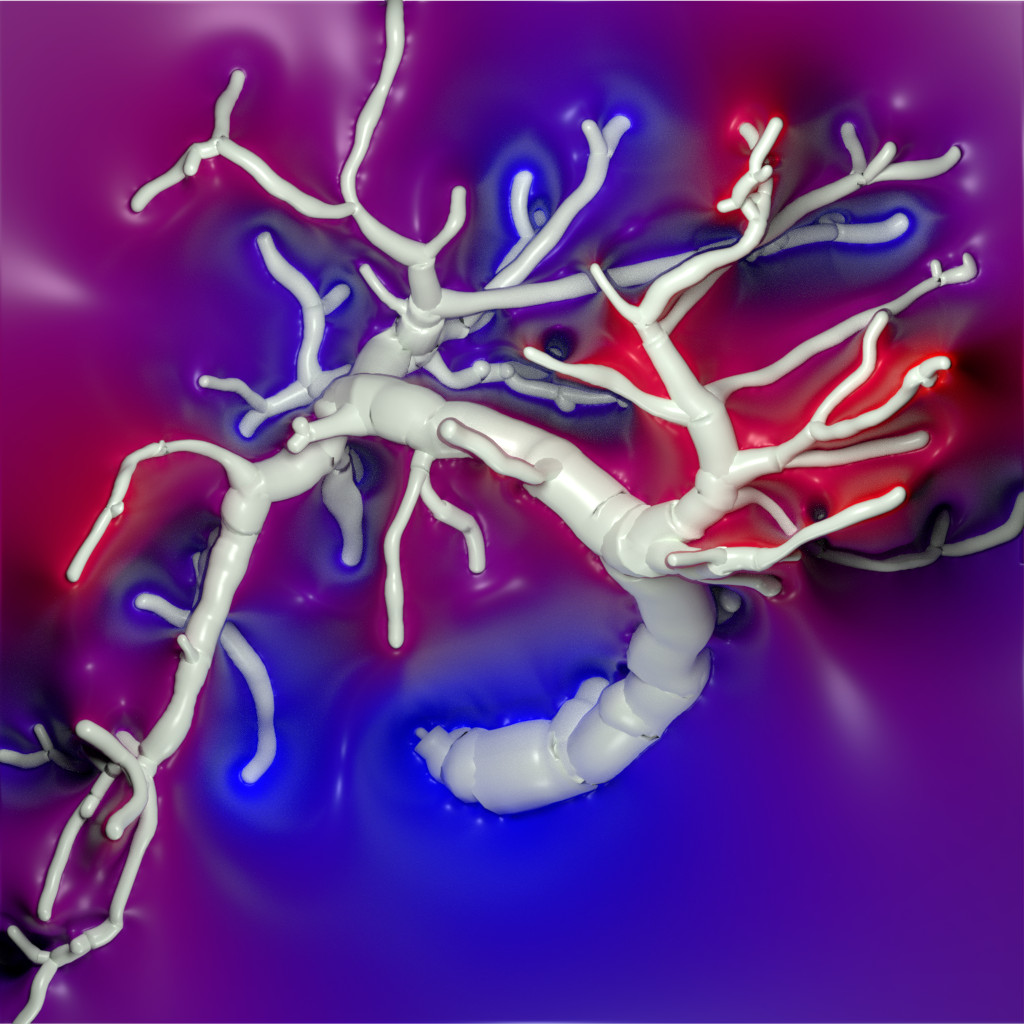}}
			\caption{}
  \end{subfigure}
  \hfill
  \begin{subfigure}[t]{0.24\linewidth}
      \centering
      \frame{\includegraphics[width=\linewidth]{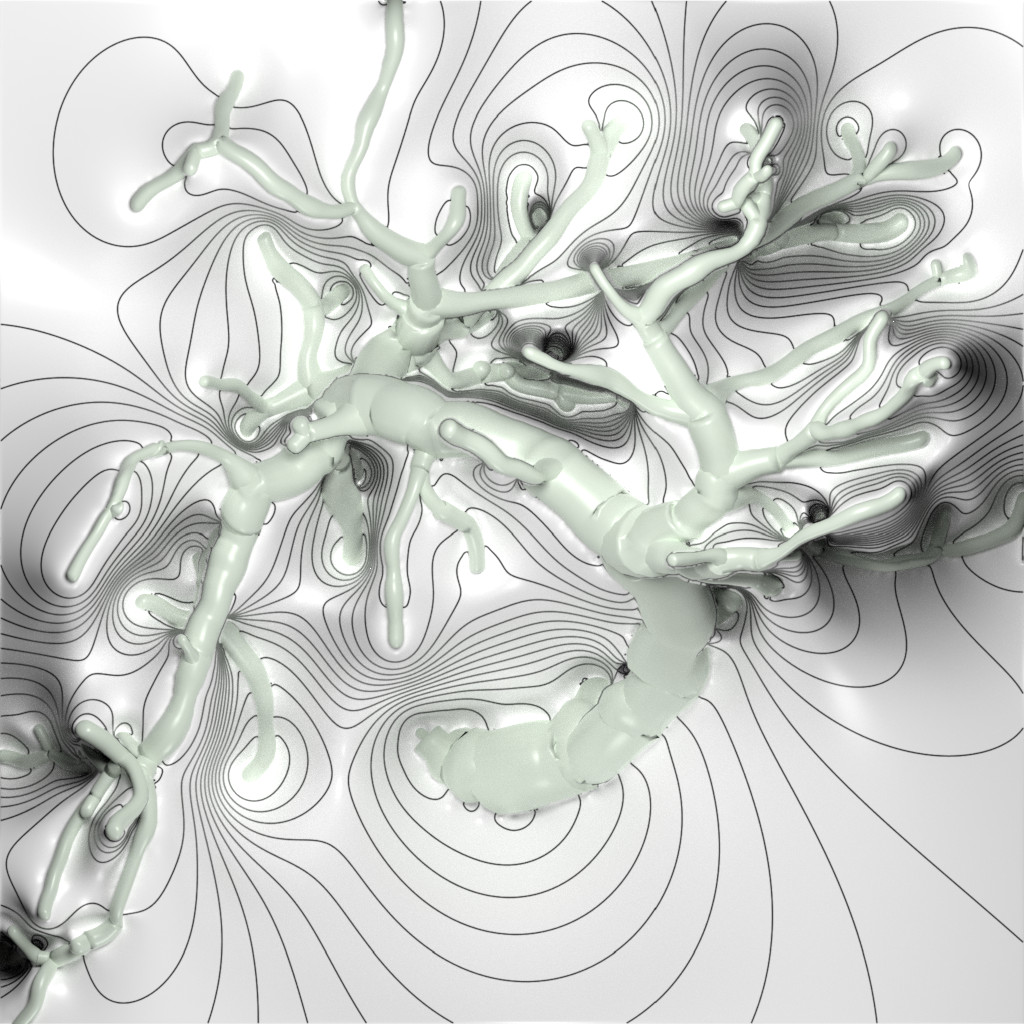}}
			\caption{}
  \end{subfigure}
  \hfill
  \begin{subfigure}[t]{0.24\linewidth}
      \centering
      \frame{\includegraphics[width=\linewidth]{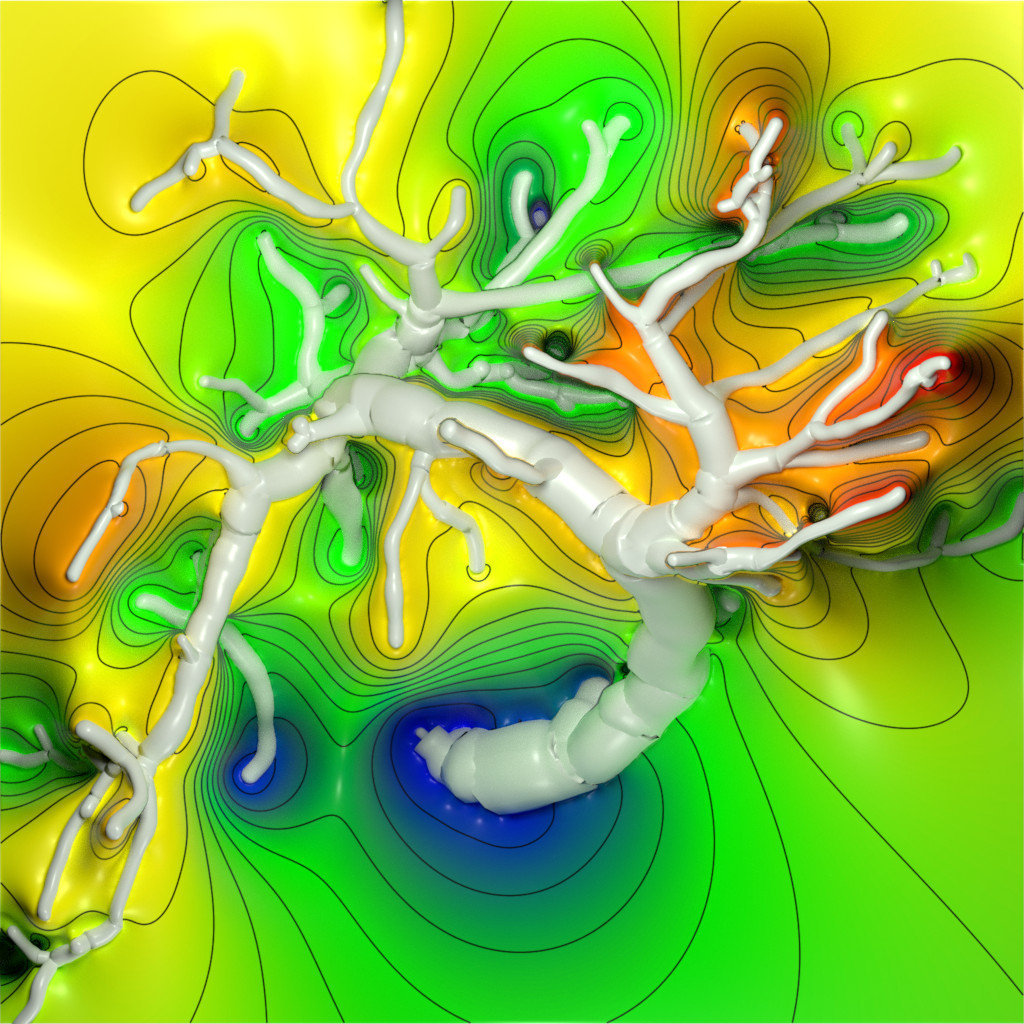}}
			\caption{}
  \end{subfigure}
  \caption{
	In order to convey depth, we incorporate different depth cues on the Void Space Surfaces.
	We encode the depth values with two different well-known color scales, chromadepth (a) and pseudo-chromadepth (b).
	We also incorporate iso-lines in our visualization (c), since they provide visual cues of the curvature and shape of the surfaces.
	Moreover, these techniques can be combined (d).
	}
  \label{fig:perceptual_enhancement}
\end{figure*}

Based on the VSS generation described in the previous subsection, we can now use the VSS as canvas for perceptual enhancements. Thus, VSS form a new foundation for a variety of depth enhancement techniques which have proven to be successful. Height field rendering would be the most similar field when it comes to the utilization of established depth and shape enhancement algorithms. In the following paragraphs, we would like to discuss a selection of depth enhancement techniques, which we found helpful in the context of vessel visualizations. We would like to point out, that this is not an exhaustive list of possibilities, but rather a subset meaningful to illustrate the capabilities of VSS. Figure~\ref{fig:perceptual_enhancement} shows the results of the perceptual enhancements using color-coding and iso-lines as well as their combination.

\noindent\textbf{Color-coding depth.} Color and brightness are commonly used to communicate depth in vessel visualization~\cite{Ropinski2006,Borkin2011,Kersten2014}. For our visualization we utilize the chromadepth and pseudo-chromadepth techniques, which exploit the effect that shorter wave lengths are refracted more than longer ones at the lens of the eye, and thus create the illusion of depth in 2D images. The full chromadepth range contains more than two colors which have no order beside their corresponding wavelength, which can be confusing for users.  Especially colors such as yellow appear brighter compared to red and can therefore wrongly be perceived closer. The reduction to pseudo-chromadepth with blue~(far) and red~(close) has the advantage that no confusion between the mental mapping from color to depth is occurring. Alternatively, the dark-means-deep metaphor can be utilized if depth values are mapped to black~(far) and white~(close) as monochromatic color scale. This supports people with any sort of color blindness while still being expressive for everyone. We apply the used colormap solely to the VSS to reserve the color channel on the vessels for parameter mappings.

%
%
%
%
%

\begin{figure}[!t]
  \begin{subfigure}[t]{0.49\linewidth}
      \centering
      \frame{\includegraphics[width=\linewidth]{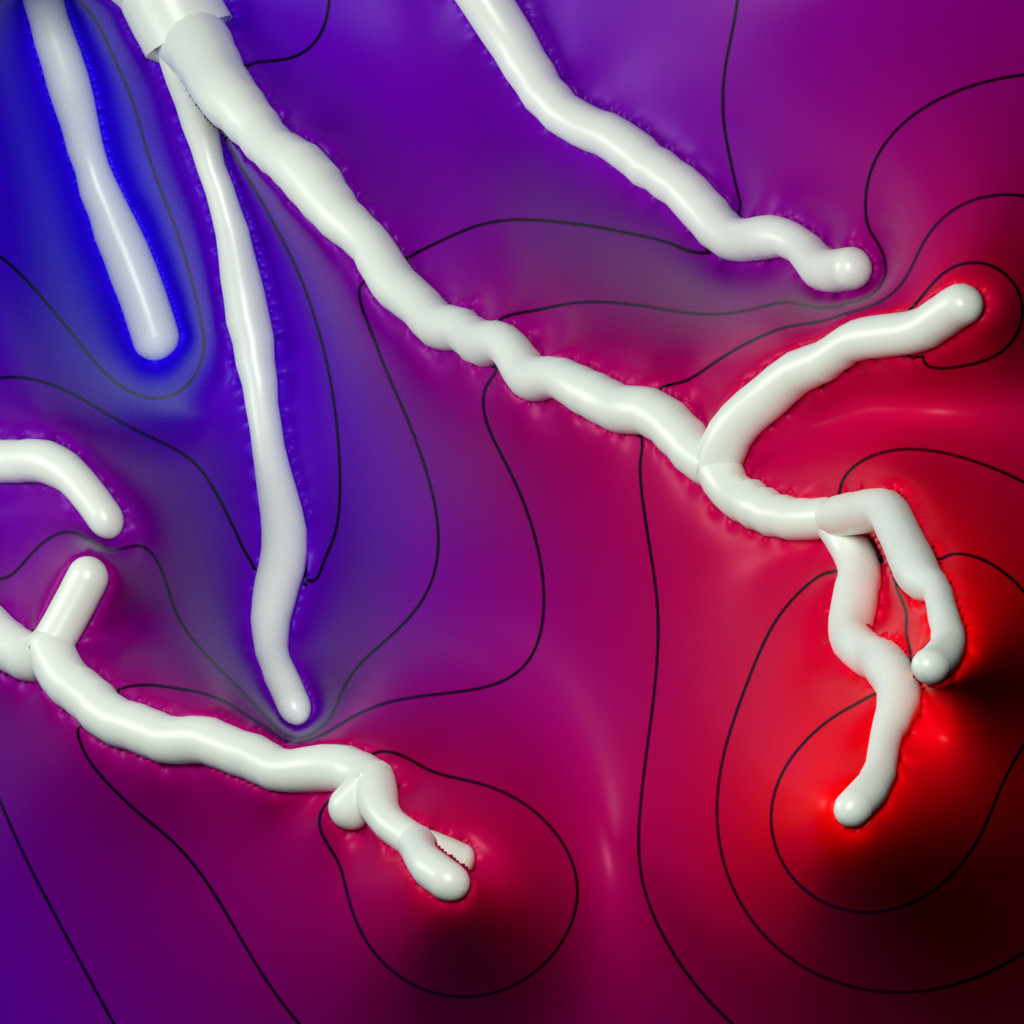}}
  \end{subfigure}
  \begin{subfigure}[t]{0.49\linewidth}
      \centering
      \frame{\includegraphics[width=\linewidth]{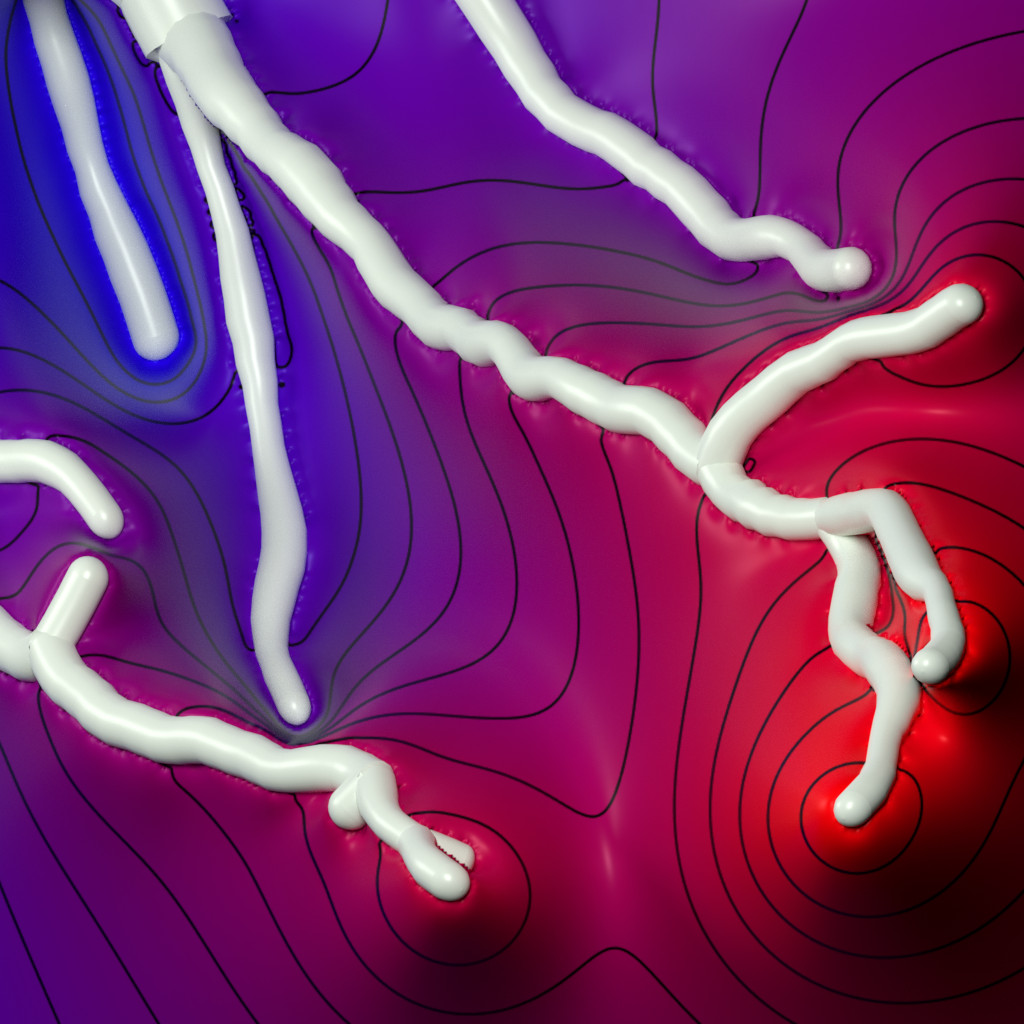}}
  \end{subfigure}
  
  \vspace{0.1\baselineskip}
  
  \begin{subfigure}[t]{0.49\linewidth}
      \centering
      \frame{\includegraphics[width=\linewidth]{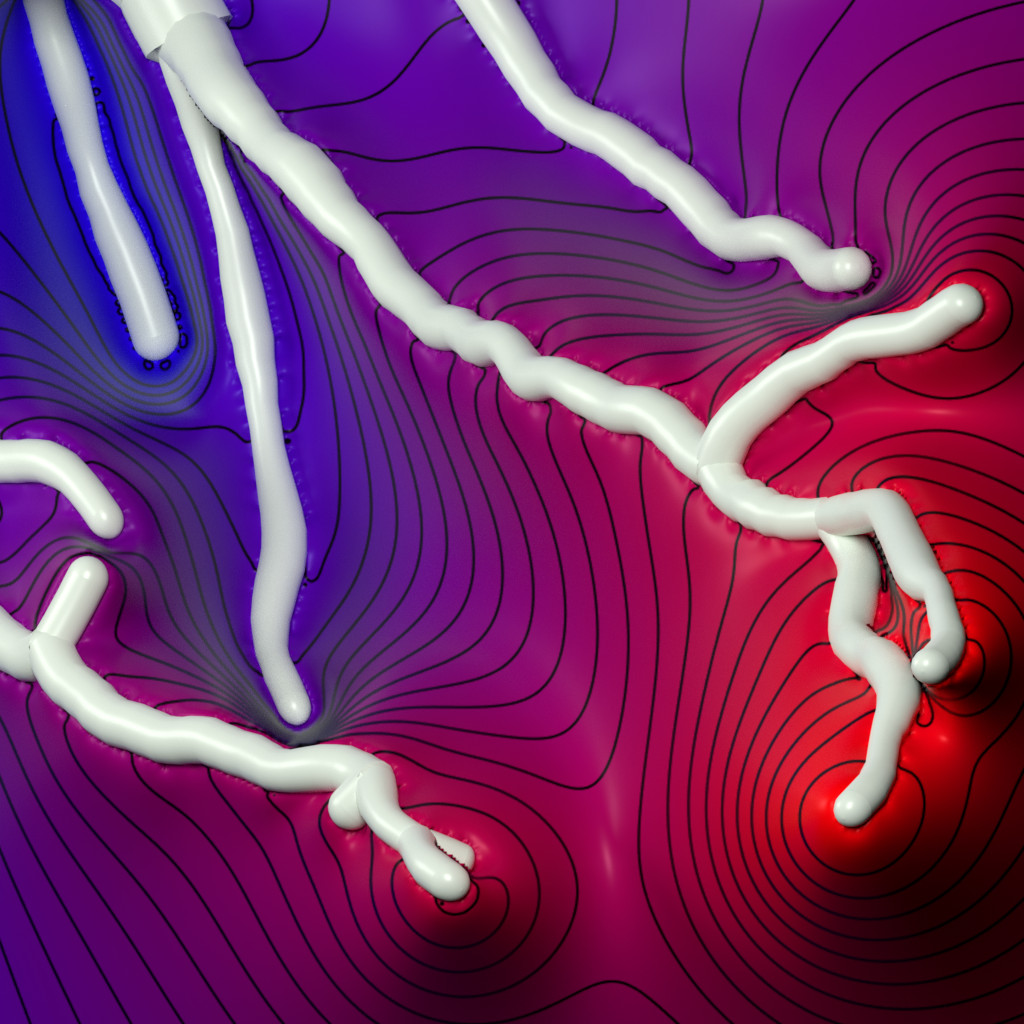}}
  \end{subfigure}
  \begin{subfigure}[t]{0.49\linewidth}
      \centering
      \frame{\includegraphics[width=\linewidth]{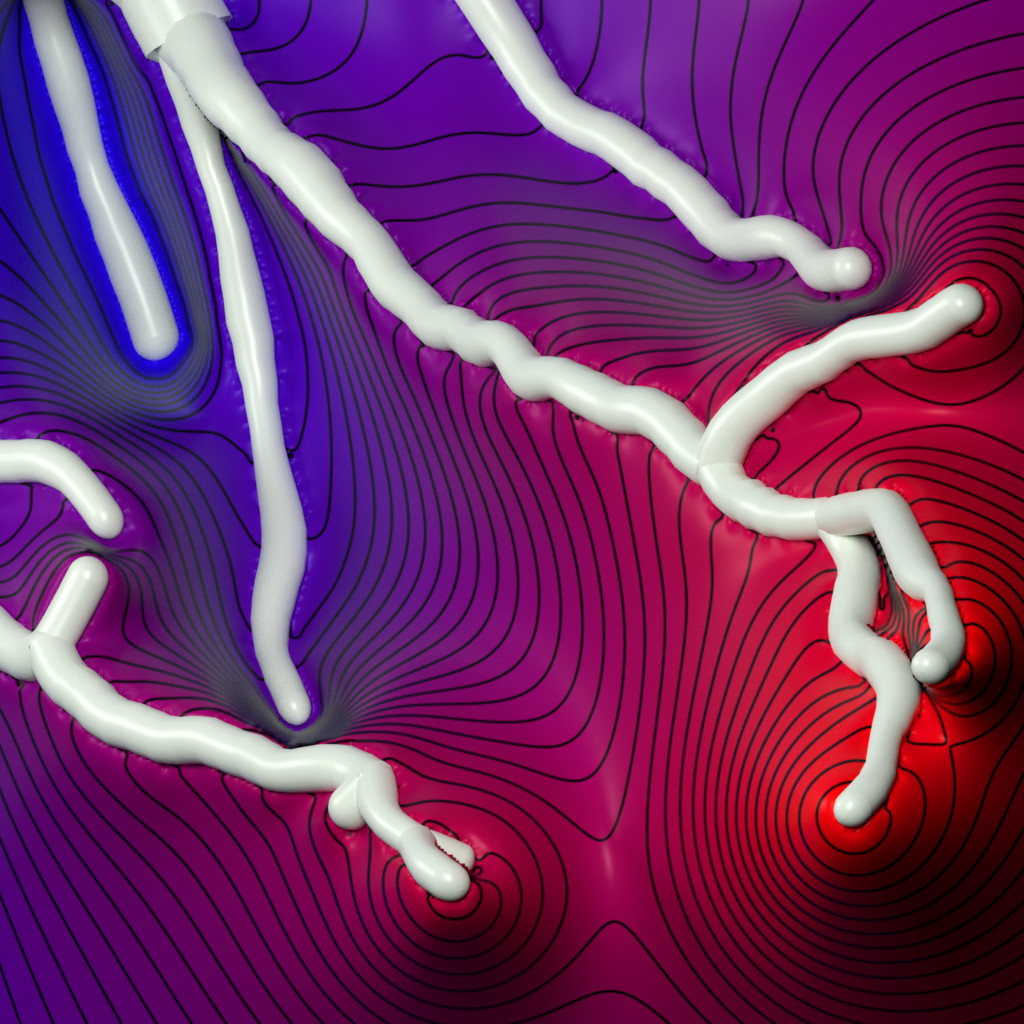}}
  \end{subfigure}
  
  \caption{
	Iso-lines describe the curvature and shape of a height field and help to convey depth and shape.
	The density of iso-lines communicates areas with a large gradient.
	These figures illustrate the effect of varying the number of iso-lines in our visualizations.}
  \label{fig:isolines_comparison}
\end{figure}

\noindent\textbf{Iso-Lines.} Inspired by topographic maps, we also incorporate iso-lines which are displayed on the VSS to aid the perception of shape. The elevation and curvature of the surface are indicated by the form and density of the level sets. More iso-lines help to convey local changes, but too many can lead to clutter at large depth differences, especially between close structures.
In the case of neighboring structures which are adjacent to the same void space, iso-lines help to distinguish points with small depth differences more easily.
A comparison of different amounts of iso-lines on our VSS is presented in Figure~\ref{fig:isolines_comparison}.

\subsection{Functional Parameter Mapping}

\begin{figure}[!t]
  \centering
  \frame{\includegraphics[width=0.49\linewidth]{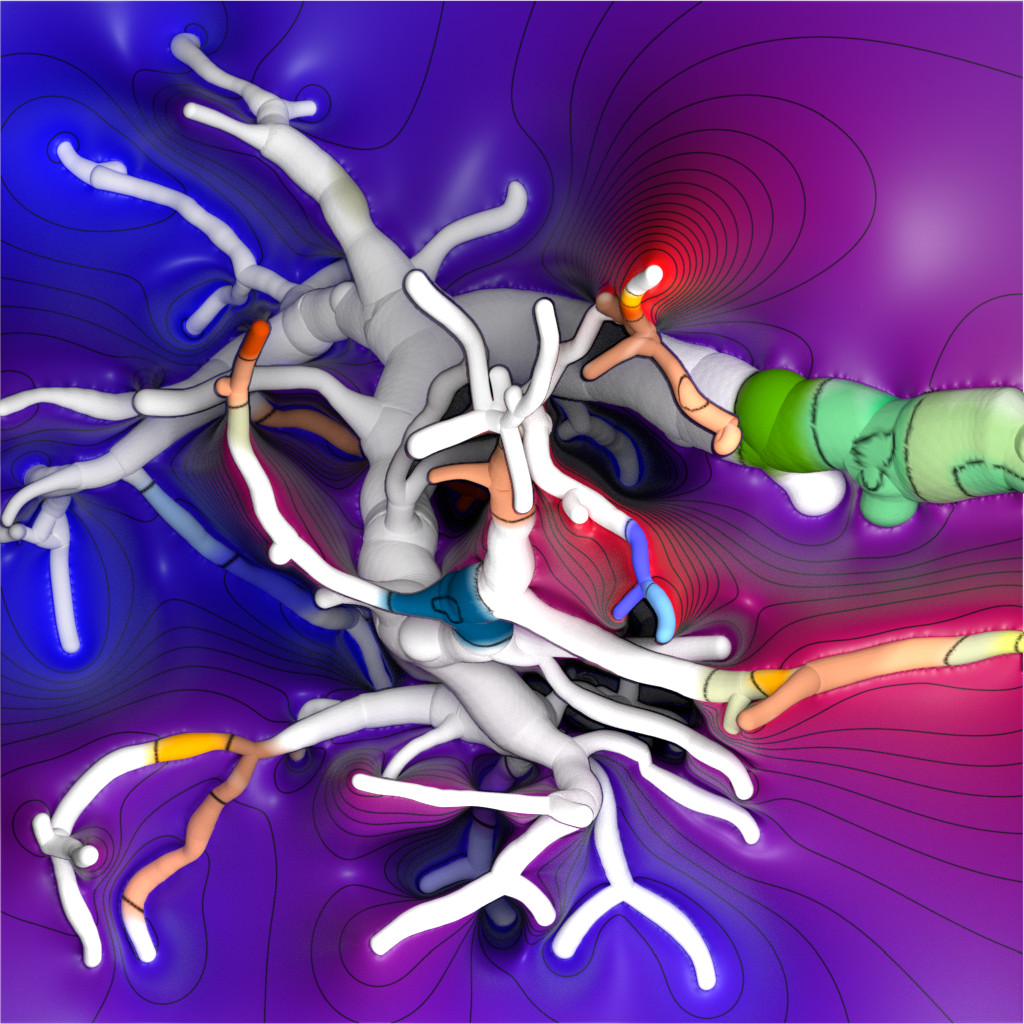}} \hfill
  \frame{\includegraphics[width=0.49\linewidth]{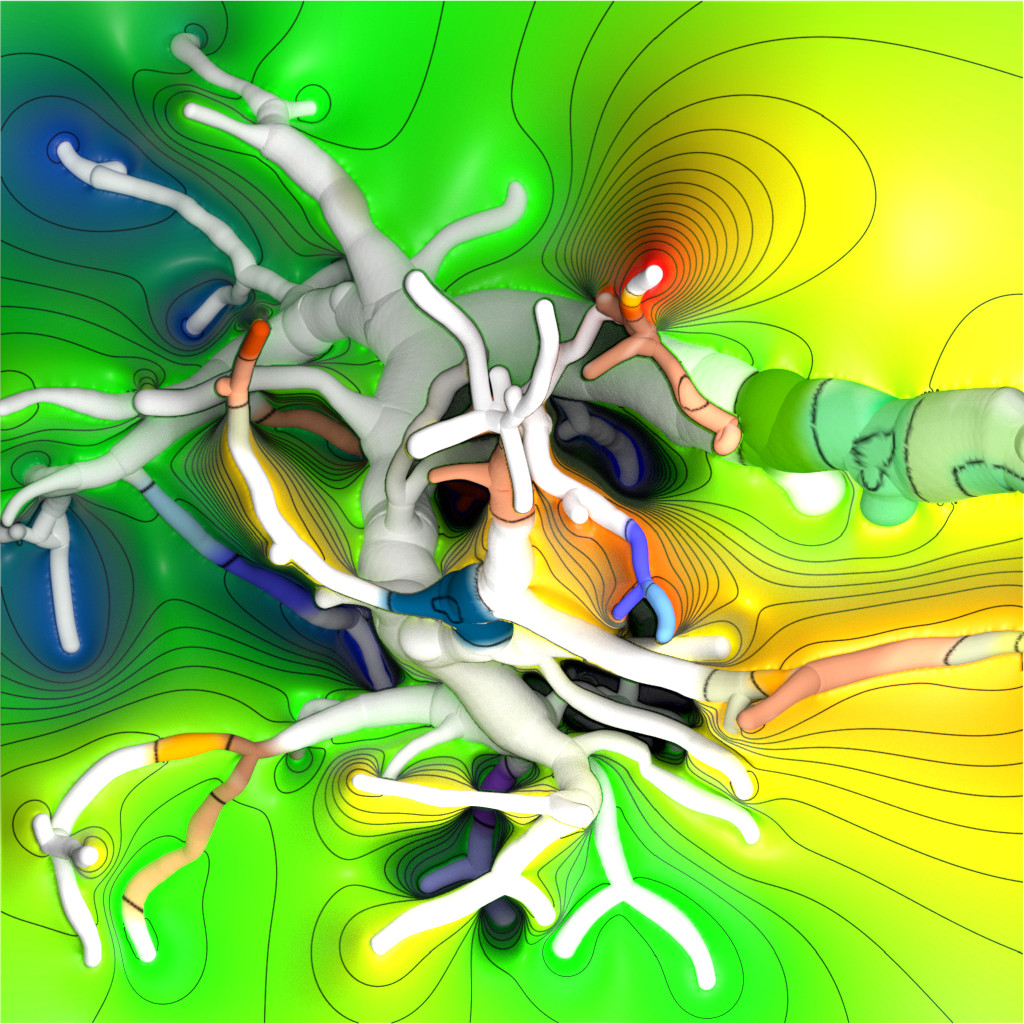}}
  \caption{Functional parameters can be mapped onto the vascular surface without direct color confusion by using VSS. These two images compare the CD and PCD color scale on our VSS with various color combinations drawn onto the vessels' geometry.}
  \label{fig:parameter_mapping}
\end{figure}

After all depth cues have been shifted away from the vascular surface model, functional parameters can be mapped on a vessel's surface without resulting in interference as shown in Figure~\ref{fig:parameter_mapping}. While communicating information such as wall shear stress or pressure, the applied color coding of these properties has to be considered. For the best contrast to the VSS, complementary or unused colors should be used instead of similar ones, which might result in confusion. Therefore, the pseudo-chromadepth color map has the advantage over the regular chromadepth range as there are already fewer colors used which could be ambiguous. The monochromatic color scheme can be used the completely eliminate this problem.

\section{Implementation Details}\label{sec:implementation}
Static 2D images of vascular structures are often used to analyze scans or to communicate results. However, inspecting the 3D model of these structures interactively can further increase the effectivity of the tasks carried out by domain experts. Therefore, we propose in this section an interactive GPU-based implementation of VSS.

\subsection{GPU Pipeline} 
The first step of our GPU pipeline is the rendering of the 3D vessel model. Since VSS only require a depth buffer as input, they are independent of the technique used to render the vessels model. Thus, both rasterization of triangular meshes or direct volume rendering, can be used in this step. Once the depth buffer is generated, our algorithm linearizes the depth values. Although, in our implementation, this is done in a separate pass, the linearization could in principle also be carried out during the rendering of the model.

In order to be consistent among different near and far plane configurations, we perform an additional pass to normalize the depth values. In this step, we determine the minimum and maximum depth values (without considering the background values), and use them to normalize all depth values to the range $[0,1]$. After this step, pixels belonging to the closest object in the screen will always have a depth value equal or close to 0 whilst pixels of the farthest object will have a depth value equal or close to 1. Within our implementation, linearized depth images are downloaded from the GPU to the CPU, where we compute the minimum and maximum depth values and also perform the normalization.

To generate the actual VSS, which interpolates the depth values on the vessel's contour, we have to distinguish between closed empty areas, the void spaces, and the pixels of their contours. To do so, we use the algorithm proposed by Suzuki et al.~\cite{suzuki1985topological}. This algorithm computes the pixels of the different contours in the image and the hierarchy between them. Thanks to this hierarchy we are able to determine the inner and outer contours of each empty area between the vessels. Moreover, we determine, for each pixel, the empty area to which they belong. In our pipeline, we used the CPU implementation of this algorithm available in the OpenCV library~\cite{opencv_library}.

The indices of the contour pixels of each void space are then uploaded to GPU memory. Moreover, we also upload to GPU memory the buffer in which, for each pixel, the identifier of the void spaces are stored. The last step of the pipeline renders a screen-aligned quad and computes the depth values for all pixels belonging to void spaces. To do so, we exploit a fragment shader, which determines the identifier of the void space adjacent to each pixel and iterates over the list of contour pixels of this space to compute the interpolated depth value by using Equation~\ref{eq:isd}. Then the interpolated depth is finally used to determine the color of the pixel based on the desired color scheme, chromadepth or pseduo-chromadepth, before we generate the iso-lines. Additionally, our fragment shader reconstructs a 3D position and normal in viewspace with which the illumination is computed.

\subsection{Performance measurements} 
We evaluated the performance of our GPU implementation on a computer with the following configuration: Intel~i7 at 3.6~GHz, with 16~GB of RAM, and a GeForce GTX980, using a screen resolution of $1280\times720$ pixels. We measured the milliseconds required to compute the different steps of our pipeline without considering the time required to render the vessel model, as this can be exchanged. The values obtained for the different configurations tested are presented in Table~\ref{tab:perf}. To investigate the effect of different void space sizes, we evaluated our algorithm with three different camera configurations: one far away from the vessels in which the branches do not intersect with the borders of the screen (column \emph{Far} on Table~\ref{tab:perf}), a camera at a medium distance in which the object intersects the screen borders only in a few places (column labeled as \emph{Medium} on Table~\ref{tab:perf}), and a close view in which the vessels intersect with the screen borders at several places (\emph{Close} column on Table~\ref{tab:perf}).
 
The results of these experiments show that the worst performance was obtained with the distant camera, which is due to two different facts. With a distant camera, the number of pixels for which the depth value has to be computed is higher than in close-up camera configurations. Moreover, in these configurations, most of the pixels of the VSS belong to the same big void space since the vessels do not intersect with the image borders. The contour of this area is composed of a high number of pixels that have to be iterated in order to properly interpolate the depth value in the VSSs, thus slowing down the performance. Table~\ref{tab:perf} also provides the maximum number of contour pixels for a single empty area and the number of contours for each visual configuration in parenthesis. As described before, the configuration in which the camera is placed far away from the object exhibits a high number of contour pixels compared with the other configurations, whilst the total number of enclosed areas is smaller.

\begin{figure}[!t]
  \centering
	\begin{subfigure}{0.49\linewidth}
      \frame{\includegraphics[width=\linewidth]{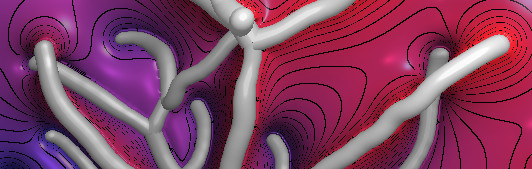}}
      \caption{Step size = 1}
  \end{subfigure}
	\hfill
	\begin{subfigure}{0.49\linewidth}
      \frame{\includegraphics[width=\linewidth]{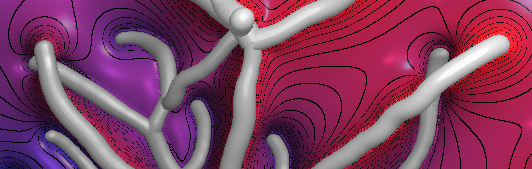}}
      \caption{Step size = 3}
  \end{subfigure}
  
	\vspace{0.5\baselineskip}  
	
	\begin{subfigure}{0.49\linewidth}
      \frame{\includegraphics[width=\linewidth]{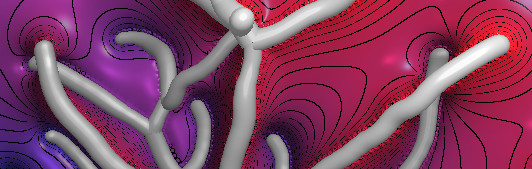}}
      \caption{Step size = 5}
  \end{subfigure}
  \hfill
	\begin{subfigure}{0.49\linewidth}
      \frame{\includegraphics[width=\linewidth]{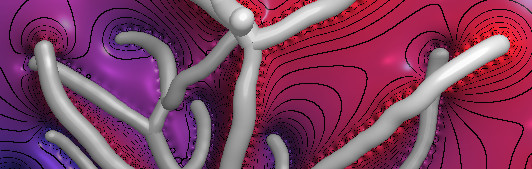}}
      \caption{Step size = 10}
  \end{subfigure}
  \caption{
	Images comparing the visual artifacts obtained when skipping contour pixels during the depth interpolation in void spaces. In these figures, a step sizes of 1, 3, 5, and 10 were applied.
	Note that even for a large step size such as 5 the visual artifacts are almost not noticeable, whilst it increases performance by more than a factor of 2.}
  \label{fig:visual_artifacts}
\end{figure}

Although we achieve interactive frame rates, when combined with other expensive rendering techniques, such as direct volume rendering and screen space ambient occlusion, we have investigated further approaches to increase the achieved frame rates. Thus, instead of iterating over all the contours pixels, the fragment shader can skip a user-defined number of pixels. As presented in Table~\ref{tab:perf}, the algorithm can be sped up by more than a factor of 2 by simply increasing the step size to 3 or 5 during the iteration of the contour pixels. However, this optimization is not free of drawbacks. Skipping a high number of pixels can introduce visual artifacts as is illustrated in Figure~\ref{fig:visual_artifacts}.

While the proposed pipeline is able to interactively generate and visualize VSS, in our current implementation, there is still room for improvement. For example, the detection of contours and their hierarchical classification could also be parallelized on the GPU, which, we believe, will even further increase the performance of VSS.

\begin{table}[!t]
	\renewcommand{\bfdefault}{bx}
  \centering
	\caption{
	In this table, the milliseconds required to generate and render the VSSs are presented.
	These values were computed for three different camera configurations: \emph{Far}, in which the object does not intersect with the borders of the screen; \emph{Medium}, in which the object intersect in a few places with the screen borders; and \emph{Close}, a close-up view in which the object intersects at several places with the screen borders. 
	Moreover, different step sizes were considered during the interpolation of the VSS's depth values.
	Lastly, we provide in parenthesis the number of independent empty areas of the image together with the maximum number of contour points among all of them.}
  \label{tab:perf}
  \begin{tabular}{l|ccc|ccc|ccc}
    \toprule
    &\multicolumn{3}{c}{Far}&
    \multicolumn{3}{c}{Medium}&
    \multicolumn{3}{c}{Close}\\
	&\multicolumn{3}{c}{(15 - 10\,K)}&
    \multicolumn{3}{c}{(17 - 5\,K)}&
    \multicolumn{3}{c}{(35 - 2.5\,K)}\\
    \midrule
    Step size&1&3&5&1&3&5&1&3&5\\
    \midrule
	\midrule
    	ms&165&114&60&
    	105&69&47&
		64&40&25\\
    \bottomrule
  \end{tabular}
\end{table}

\section{Evaluation}\label{sec:userstudy}
To evaluate the perceptual performance of VSS, we conducted a lab study to compare against other established vessel visualization techniques.
In particular, we were interested in how fast and accurate participants can judge depth using different kinds of cues.
The approaches we compared against each other were three different 3D visualization techniques -- two of them in two variations which makes a total of five scenarios.
We designed the comparative study as within-subject, whereby each variation of the five visualizations was presented to every participant.

The first pair consists of a neutral vessel material and our VSS with iso-lines and global illumination, one with chromadepth~(VSS-CD) and the other using the pseudo-chromadepth~(VSS-PCD) color mapping.
The second pair applies the chromadepth~(vessel-CD) and pseudo chromadepth~(vessel-PCD) color scale using global illumination directly onto the vascular surface as introduced by Ropinski et al.~\cite{Ropinski2006}.
Finally, we utilize a neutral vessel rendering~(baseline) using global illumination without additional depth cues as baseline reference to the first two pairs of enhanced visualizations.
The labeling for each of the five configurations in parentheses is used in all further descriptions and figures as short notation.


\subsection{Hypotheses}
To evaluate participants ability to make fast, and therefore intuitive judgments on a given task, we decided to measure response times as well as accuracy.
Therefore, we have formulated four hypotheses about what we expect from the compared visualization setups.

\begin{enumerate}[noitemsep,topsep=5pt,parsep=2pt,partopsep=0pt]
	\item The baseline performs worst in both aspects~(response time and accuracy).
	\item Due to the indirectness, the response time if using VSS is slower compared to depth cues directly rendered on the vessel's surface.
	\item VSS and the direct vessel rendering perform similar in terms of accuracy.
	\item Based on the findings by Ropinski et al.~\cite{Ropinski2006}, we hypothesize that the pseudo-chromadepth color mapping of both setups, with and without the VSS, is superior to their chromadepth version in both aspects~(response time and accuracy).
\end{enumerate}

\subsection{Stimuli and Task}\label{subsec:stimuli}
\begin{figure}[!t]
  \begin{subfigure}[t]{0.49\linewidth}
      \centering
      \frame{\includegraphics[width=\linewidth]{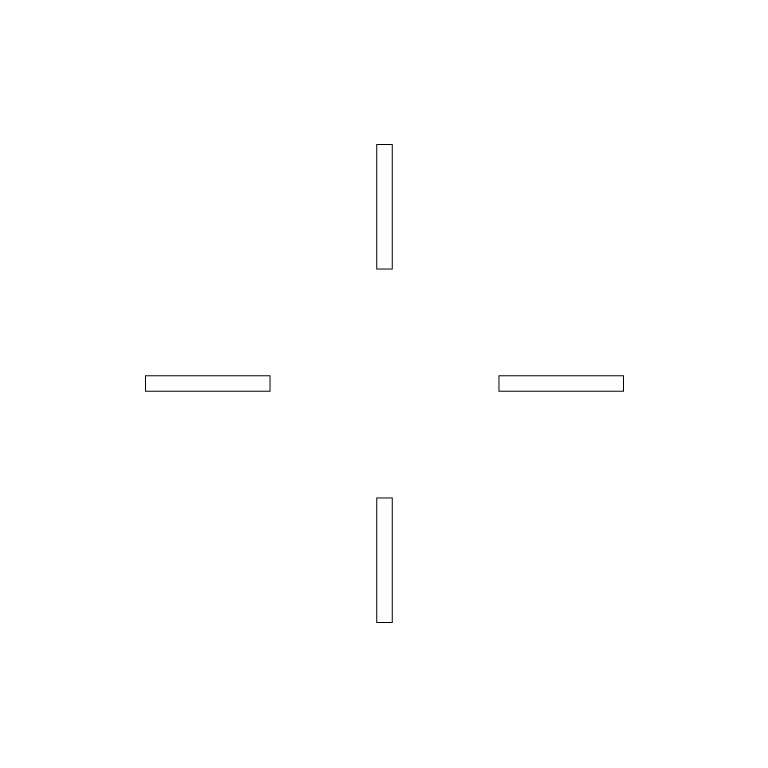}}
      \caption{}
  \end{subfigure}
  \begin{subfigure}[t]{0.49\linewidth}
      \centering
      \frame{\includegraphics[width=\linewidth]{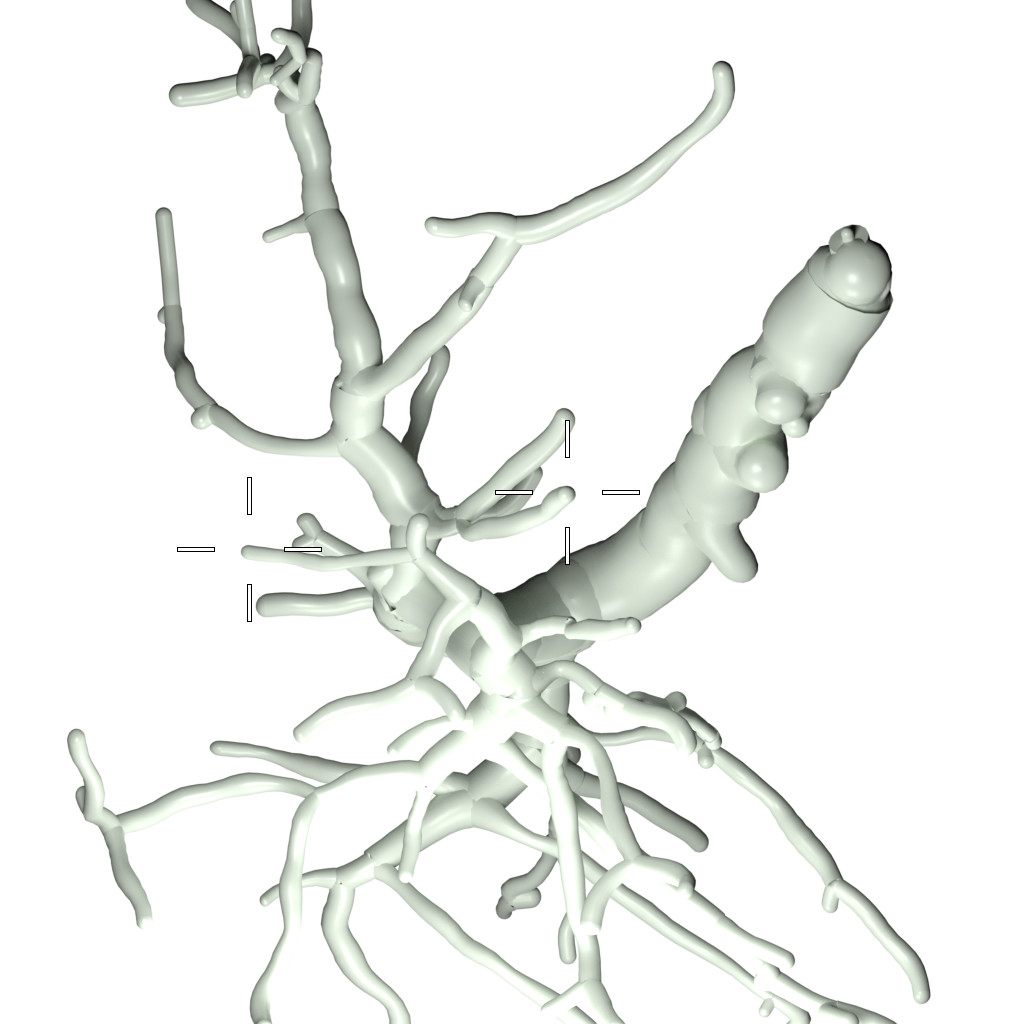}}
      \caption{}
  \end{subfigure}
  
  \vspace{0.5\baselineskip}
  
  \begin{subfigure}[t]{0.49\linewidth}
      \centering
      \frame{\includegraphics[width=\linewidth]{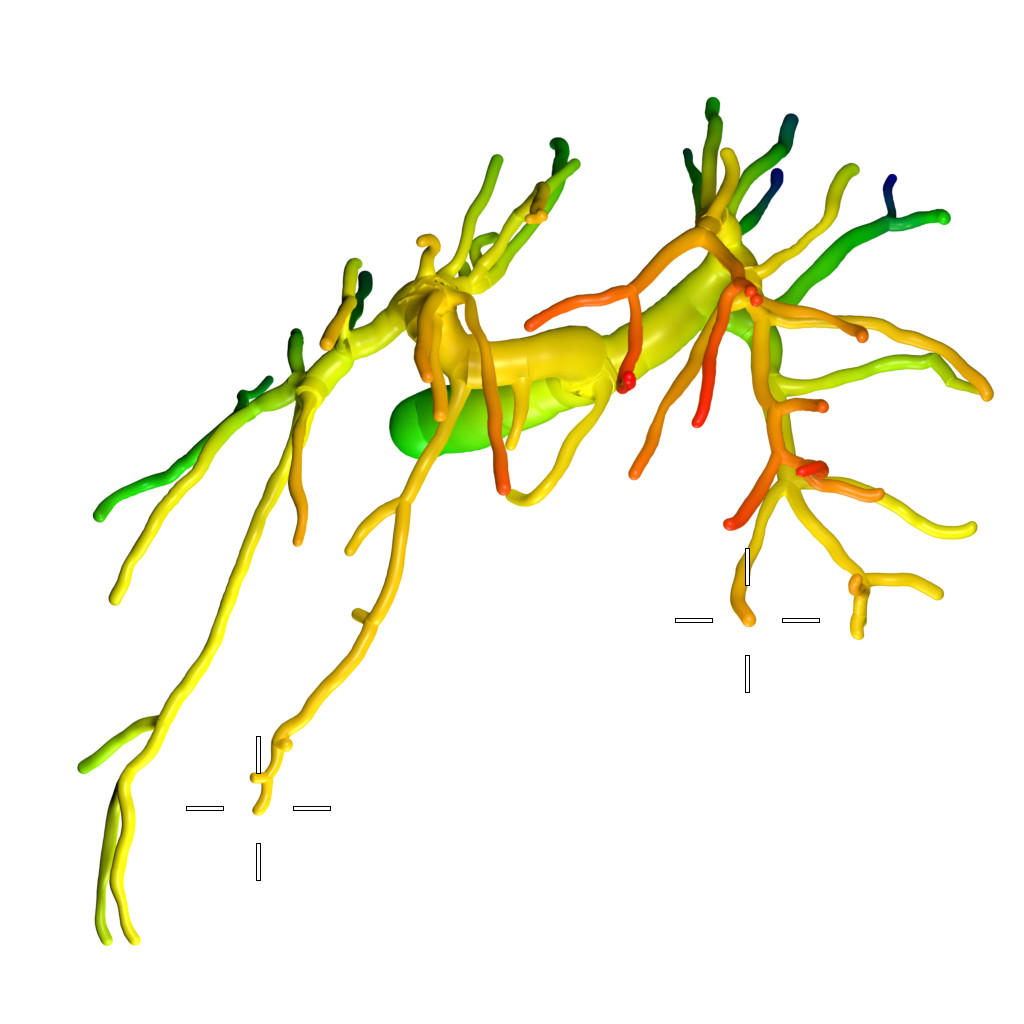}}
      \caption{}
  \end{subfigure}
  \begin{subfigure}[t]{0.49\linewidth}
      \centering
      \frame{\includegraphics[width=\linewidth]{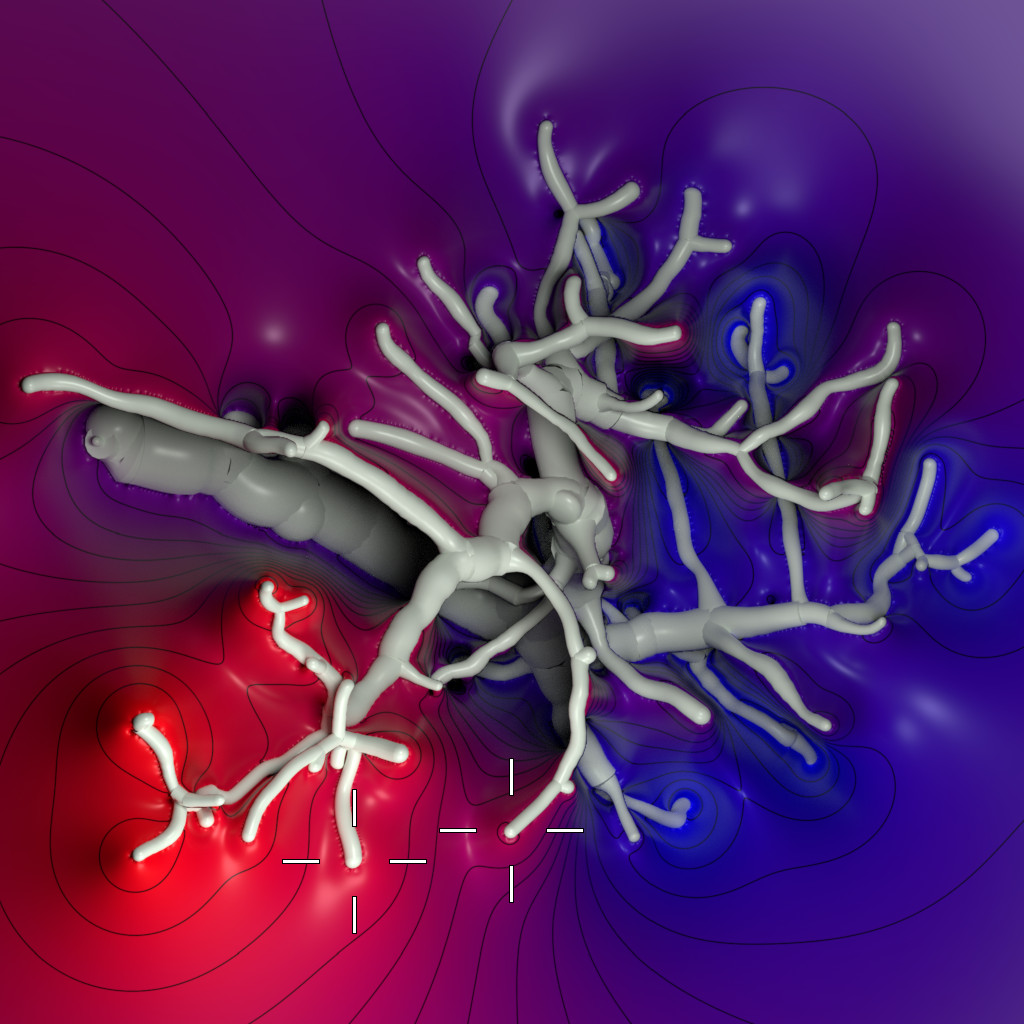}}
      \caption{}
  \end{subfigure}
  
  \caption{
	In our user study, we presented to the participants several figures generated using different configurations, such as plain shading~(b), colored vessel~(c), or colored VSS~(d).
	Two points of the vessel structure were highlighted by using the marker in Figure~(a).
	The participants had to select the closest point to the camera between them.}
  \label{fig:marker_and_stimuli}
\end{figure}

When designing our study we kept the task complexity to a minimum, and tried to reduce the amount of cognitive load for the participants.
To achieve this, each image presented to the participant consists only of a vascular structure and two markers pointing on the model's surface.
Figure~\ref{fig:marker_and_stimuli} shows an example of our used marker design and three different stimuli.
The task was then to decide whether the left or right position appears closer when compared to the other marker.
For that purpose, we designed the markers to be as less distracting as possible while still pointing clearly to the point of interest.
As a result, the search time for both markers in the image was minimized before the actual decision can be done by the participant.
Therefore, we utilize a cross with four separated and spread out arms.
This shape avoids too much of occlusion of the actual target while still standing out due to its regular shape compared to the organic vessel appearance.
The interior of the cross is colored white and the border black to achieve high contrast, independent of the structure and hue below.

To balance the task complexity per displayed image, we arranged the pair of markers with three different criteria.
Their distance in screen space, target depth difference, and if they are directly connected through the void space.
We assured that the markers are at least~10\% of the total depth difference in the image apart and clearly separated into left and right.

Using our visualization techniques~(5), different vascular models~(6), and different positions of markers~(5), we create a total of 150 stimuli.

\subsection{Participants}
\newcommand{\nummaleparticipants}{15}
\newcommand{\numfemaleparticipants}{5}
\newcommand{\numtotalparticipants}{21}
\newcommand{\numvalidparticipants}{20}
\newcommand{\numoutlierparticipants}{one}
\newcommand{\participantsagemin}{22}
\newcommand{\participantsagemax}{34}
\newcommand{\participantsagemean}{30}
\newcommand{\participantsagestddev}{3.1}

We conducted an evaluation with~\numvalidparticipants{} participants from our university campus population (\numfemaleparticipants{} female, \nummaleparticipants{} male), aged \participantsagemin{}~-~\participantsagemax{} ($M = \participantsagemean{}$, $SD = \participantsagestddev{}$).
We excluded one participant from further analysis, because of error rates at $\approx$ 50\%, indicating that the given task was not understood well enough.
All participants had normal or corrected to normal vision and none of them had any further impairments such as color blindness.

\subsection{Procedure}
In the beginning of the study, we briefly introduced the participant into the topic of depth and shape perception.
Afterwards, the controls of our application were explained.
We specified a left~('F') and right~('J') key corresponding to the left and right marker respectively.
Participants were instructed to give an answer as quickly as possible, while still making sure to be certain about which point appears to be closer to them.

Each stimuli was started with a focus screen showing a countdown for three seconds, before presenting the stimuli.
After a response was given, participants had the chance to focus again, before moving to the next stimuli by pressing the space bar.
This neutralizes the previously displayed stimulus and assures that the participant is prepared to give an answer.

Participants started the survey by completing a training phase to familiarize with the task.
Answers in this phase could be given until the participants feels familiar with the process.
Afterwards, the main evaluation was presented consisting of our stimuli as described in Section~\ref{subsec:stimuli}.
The set of images with all configurations was randomized and counterbalanced using Latin square.
Each of the images includes two markers, clearly highlighting a point on a vessel.

\subsection{Statistical Analysis}
Since our study is designed as within-subject and our data is non-parametric~(verified by the Shapiro-Wilk test for normality), we used the Friedman's ANOVA~(analysis of variance).
We found significant differences between the compared systems for response times~($\chi^2(4)~=~49.024$, $p~<~.001$), as well as for accuracy~($\chi^2(4)~=~43.68$, $p~<~.001$).
An additional post hoc analysis with Bonferroni correction for both measured aspects~(response time and accuracy) exhibits all pairs of visualization configurations where a significant difference exists.

\begin{figure}[!t]
  \centering
  \begin{subfigure}[t]{0.49\linewidth}
      \centering
      \includegraphics[width=\linewidth]{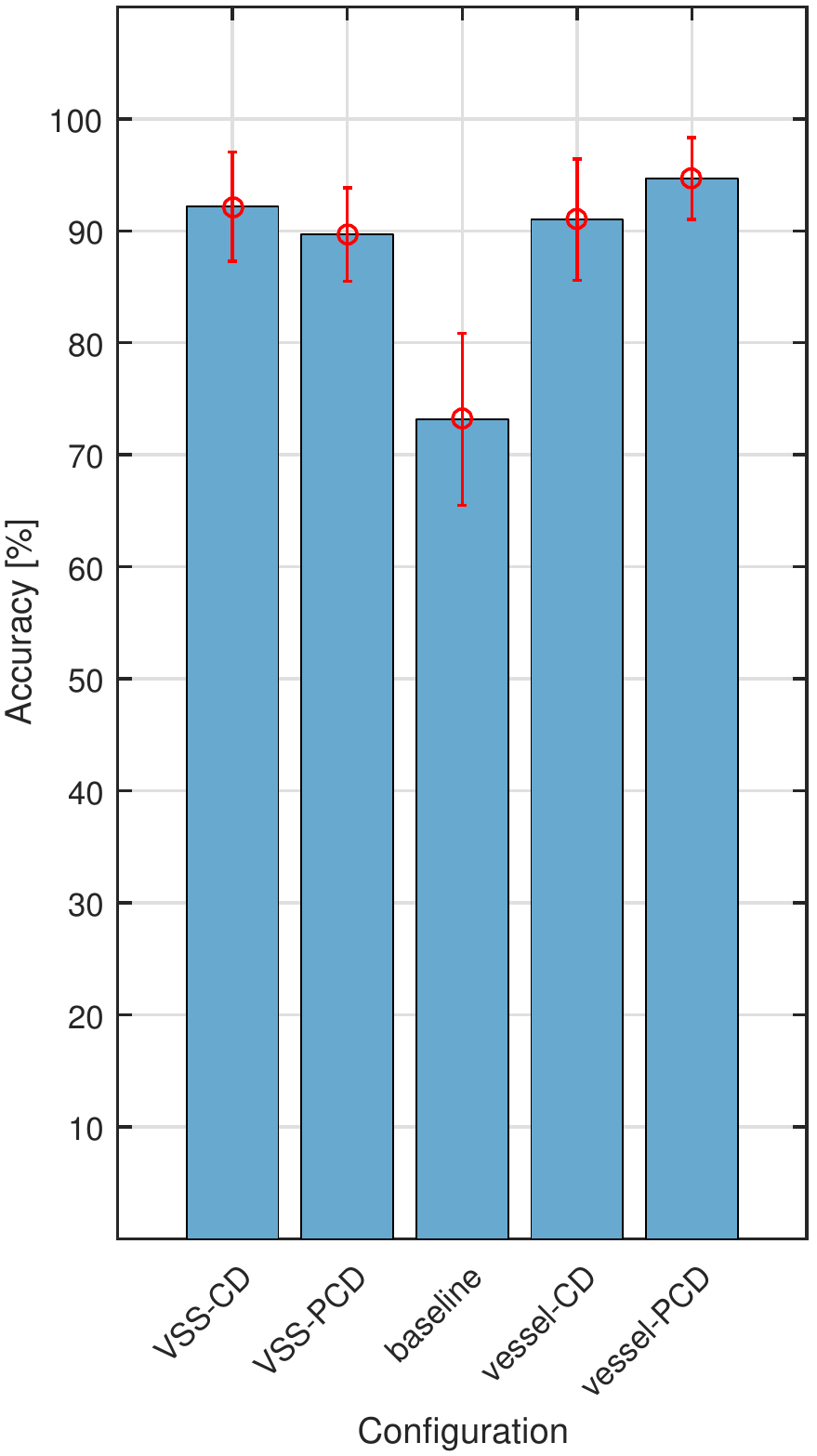}
      \caption{Error bar chart for accuracy in percent per visualization system.}
      \label{fig:boxplot_barchart:accuracy}
  \end{subfigure}
  \hfill
  \begin{subfigure}[t]{0.478\linewidth}
      \centering
      \includegraphics[width=\linewidth]{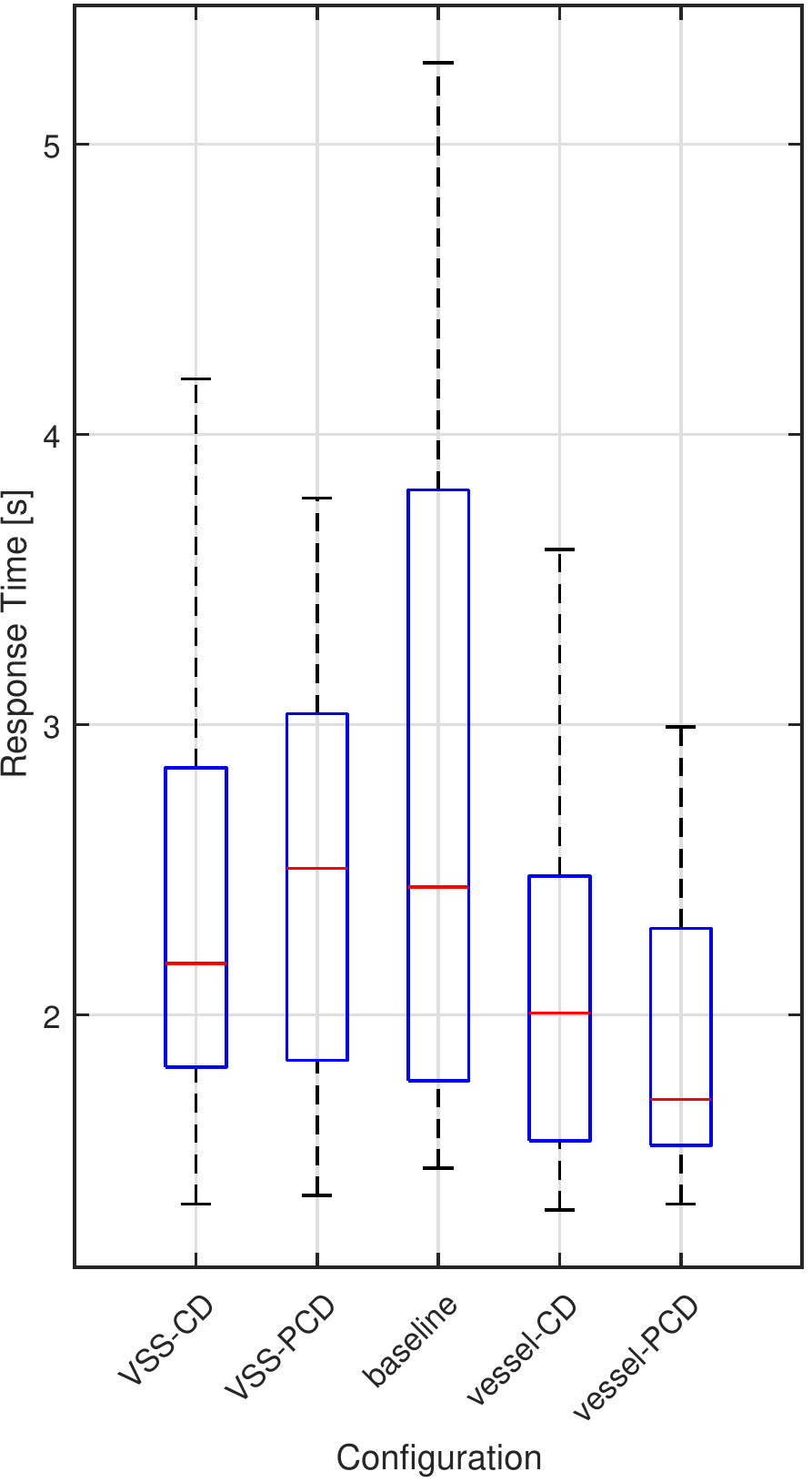}
      \caption{Box plot for response time in seconds per visualization system.}
      \label{fig:boxplot_barchart:time}
  \end{subfigure}
  
  \caption{For each participant in our study, we measured the accuracy~(a) and response time~(b) during a depth judgment task. The here shown statistics represent 20 participants with each of them assessing 150~stimuli.}
  \label{fig:boxplot_barchart}
\end{figure}

All of the depth enhanced visualization performed similar and outperformed the neutral baseline rendering in terms of accuracy.
In all cases the critical difference was 28.07034 with observed differences of VSS-CD~$=$~50.5, VSS-PCD~$=$~39.5, vessel-CD~$=$~46.0, and vessel-PCD~$=$~64.0. Error bar charts for each of the technique are shown in Figure~\ref{fig:boxplot_barchart:accuracy}.

Regarding response times, we found significant differences between the methods which use depth cues directly on the vascular surface~(vessel-CD and vessel-PCD) and all three other configurations~(VSS-CD, VSS-PCD, and the baseline).
The critical difference is the same as for the accuracy measurements with 28.07034.
The pairwise observed differences are VSS-CD~$\leftrightarrow$~vessel-CD~=~34.0, VSS-CD~$\leftrightarrow$~vessel-PCD~=~48.0, VSS-PCD~$\leftrightarrow$~vessel-CD~=~34.0, VSS-PCD~$\leftrightarrow$~vessel-PCD~=~48.0, baseline~$\leftrightarrow$~vessel-CD~=~36.0, and baseline~$\leftrightarrow$~vessel-PCD~=~50.0. A box plot presenting all timing measurements is shown in Figure~\ref{fig:boxplot_barchart:time}.



\section{Discussion and Comparison}\label{sec:discussion}
In this section we discuss the results obtained in our user study and the benefits of our method as compared to the state of the art.
Moreover, at the end of this section, we detail the limitations of our technique.

\subsection{Study results}
The results obtained in our user study supports most of our hypothesis.
Our first hypothesis stated that the baseline method performs worst in response time and accuracy than when providing depth cues on the vessels' surface or on our VSS.
The results of the user study support that both advanced techniques performed better than the baseline in terms of accuracy.
Regarding the response time, however, that was only true for the case in which depth cues were applied directly on the vessels' surface.
We believe that this is due to the indirection introduced by our method since the user needs to also focus their attention on the surrounding areas of the vessels.

Our results also support Hypothesis 2.
As in the previous case, the response time obtained with our method was significantly lower than the one achieved by applying depth cues directly on the vessels' surface.
As mentioned before, in order to perceive the depth cues with our method the user needs to focus its attention on the surrounding areas of the vessel, thus increasing the time until a user can judge depth.

Our third hypothesis predicted that both techniques, depth cues on the vessels and depth cues on the VSSs, would achieve similar accuracy which our results also supports.
Although, our method did not improve the existing technique in terms of accuracy, it achieved a similar performance whilst leaving room on the vessels' surface to communicate additional information.
Contrary to the study by Ropinski et al.~\cite{Ropinski2006}, our study could not provide support on the evidence that the pseudo-chromadepth color mapping is superior over the full range chromadepth technique.

Moreover, the participants of our user study presented a similar type of response to our images.
Between VSS and vessel surface coloring, the last one shown up as the most effective in terms of speed indicating that participants where able to make an intuitive judgement.
However, with VSS, iso-lines were perceived as very helpful in the configurations in which the depth difference between the queried points was small and the points are adjacent to the same void space.
However, if the points do not share a void space, color coding was the most effective depth cue, which underlines the need for combination of depth cues.
Although the color range of the pseudo-chromadepth color mapping is reduced, it was perceived as most helpful.
We believe that the high performance obtained with the pseudo-chromadepth method is due to the fact that a linear relationship can be established between depth and color.
The user can associate blue with far objects and red with close ones, something not possible with the chromadepth color scale.
Furthermore, some participants mentioned that the chromadepth technique was confusing since they associated yellow with close objects due to its brighter appearance.
Lastly, the participants of our study sorted the techniques from more useful to less useful as follows: pseudo-chromadepth on the vessels' surface, pseudo-chromadepth on the VSSs, chromadepth on the vessels' surface, chromadepth on the VSSs, and regular 3D rendering.

In the future, we plan to conduct further studies to investigate which depth cue has the biggest effect on the depth perception.
We plan to find this our by determining which is the minimum depth a user is able to differentiate while using different techniques, and which combination of parameters performs best.

\subsection{State of the Art Comparison}
\begin{figure}[!t]
  \begin{subfigure}[t]{\linewidth}
      \centering
      \frame{\includegraphics[width=0.49\linewidth]{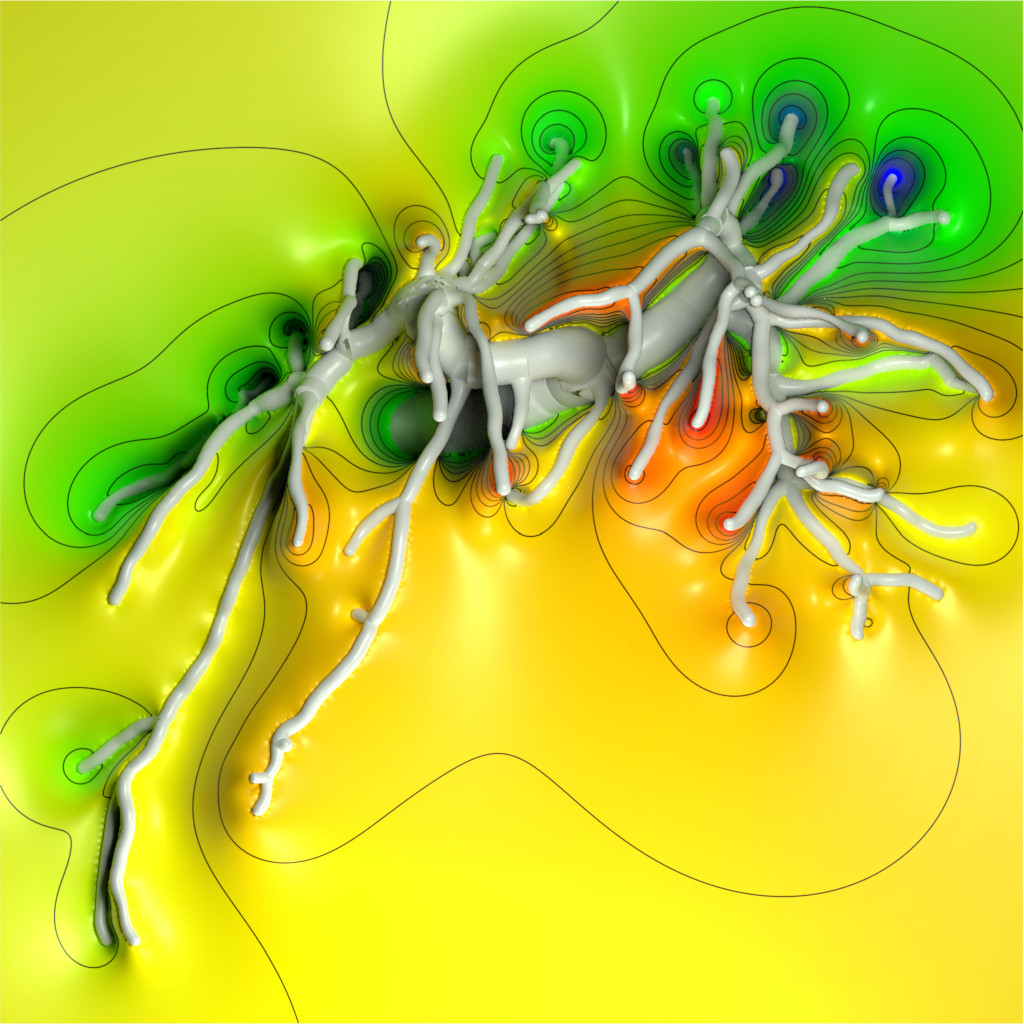}}\hfill
      \frame{\includegraphics[width=0.49\linewidth]{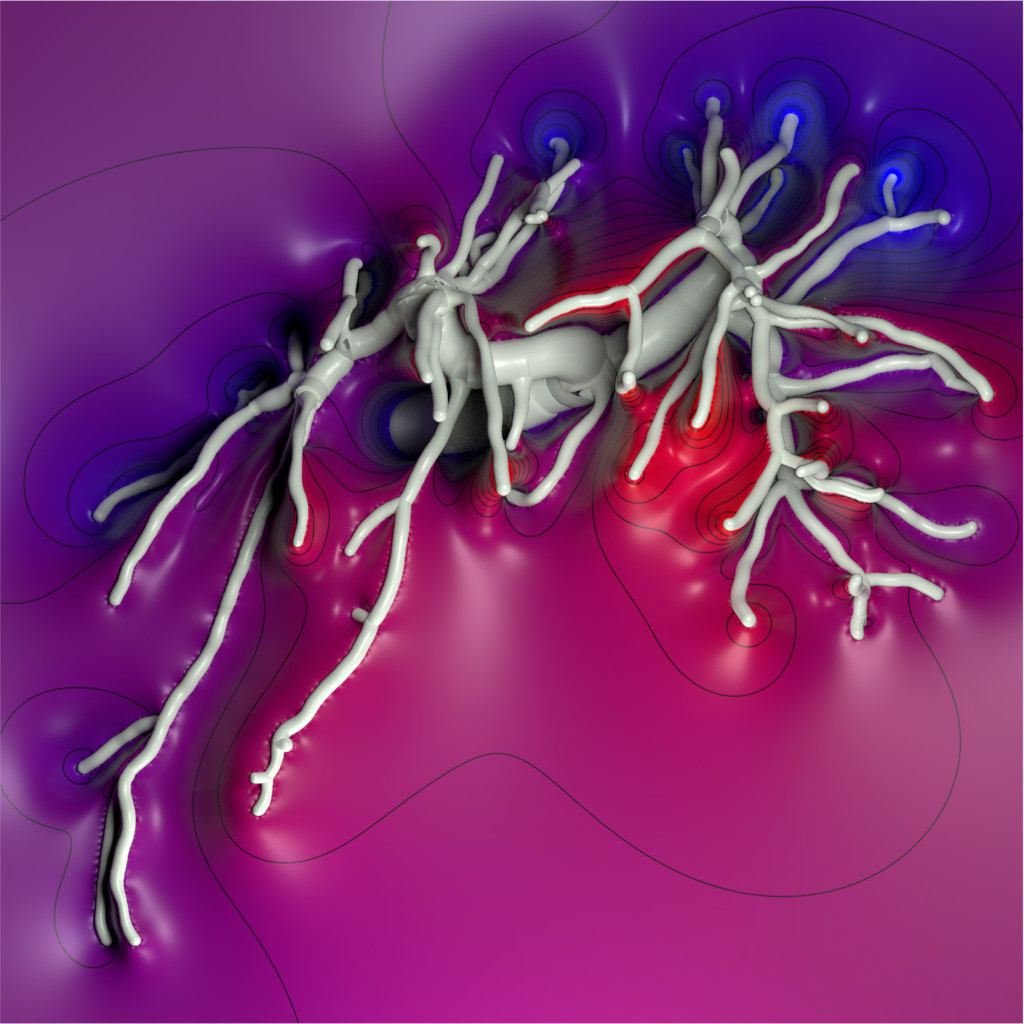}}
  \end{subfigure}
  
  \vspace{0.5\baselineskip}
  
  \begin{subfigure}[t]{\linewidth}
      \centering
      \frame{\includegraphics[width=0.49\linewidth]{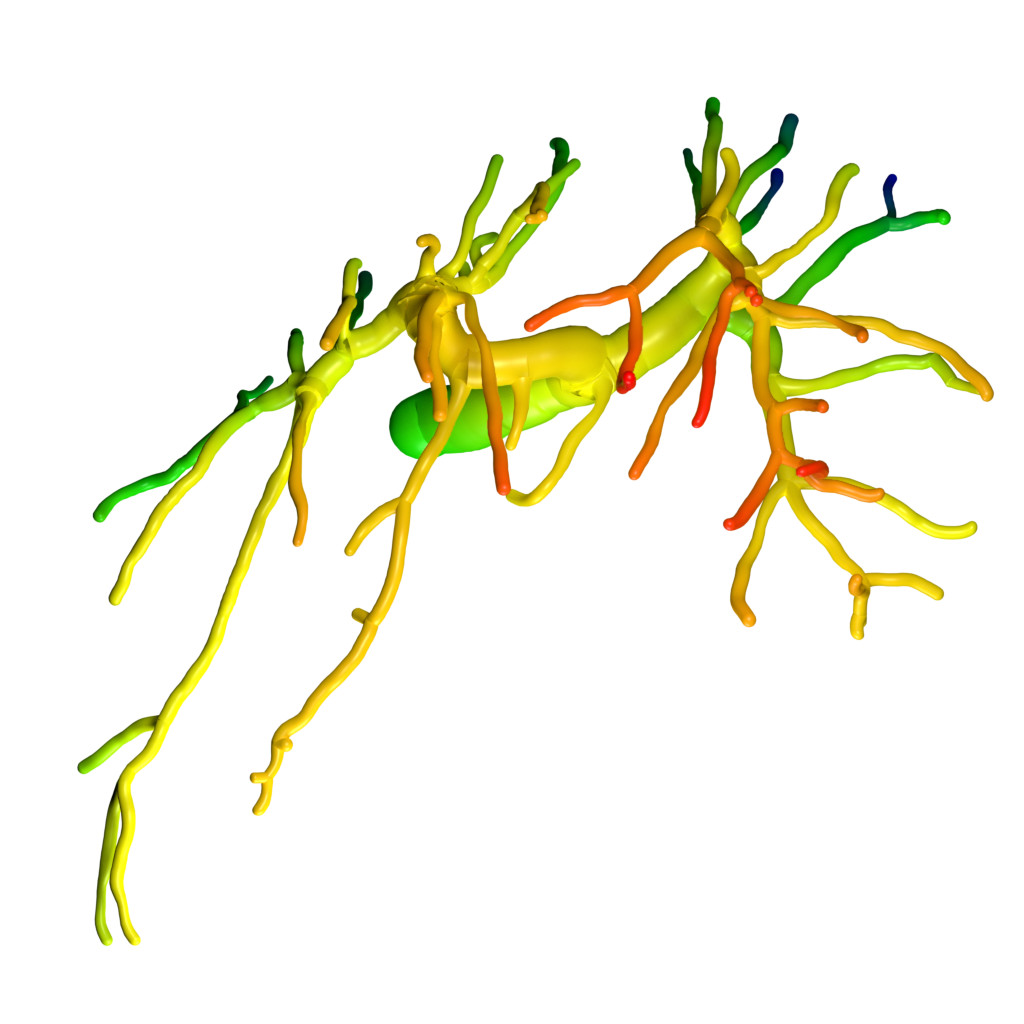}}\hfill
      \frame{\includegraphics[width=0.49\linewidth]{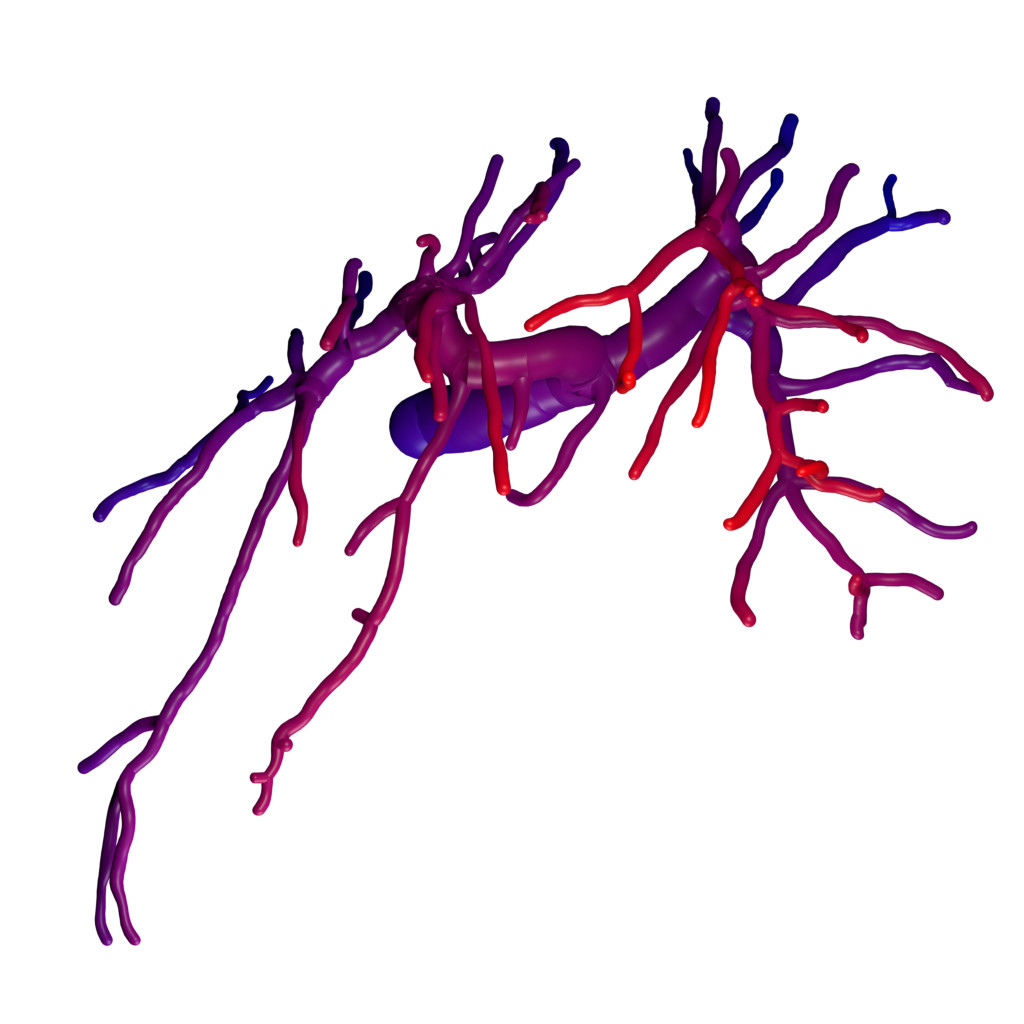}}
  \end{subfigure}
  
  \caption{
	Comparison between our method (top figures) and the chromdepth (bottom left) and pseudo-chromadepth~\cite{Ropinski2006} (bottom right) methods.}
  \label{fig:comparison_ropinski}
\end{figure}

Our user study takes into account the enhancement of depth perception from Ropinski et al.~\cite{Ropinski2006} using the chromadepth and pseudo-chromadepth color scheme directly on the vessel tree as shown in Figure~\ref{fig:comparison_ropinski}.

We also compared our technique with the method proposed by Behrendt et al.~\cite{Behrendt2017}, which is to our knowledge the only method which allows for the visualization of additional parameters while depicting depth cues in place without using symbols. In contrast to VSS, their technique uses the vessels structures to communicate depth and additional parameters. They convey depth through the pseudo-chromadepth color scale encoded on the areas close to the edges of the vessels in a similar way as the Fresnel equations compute the reflection term. The central areas of the vessels then are used to encode other parameters.

\begin{figure}[!t]
  \centering
  \frame{\includegraphics[width=0.49\linewidth]{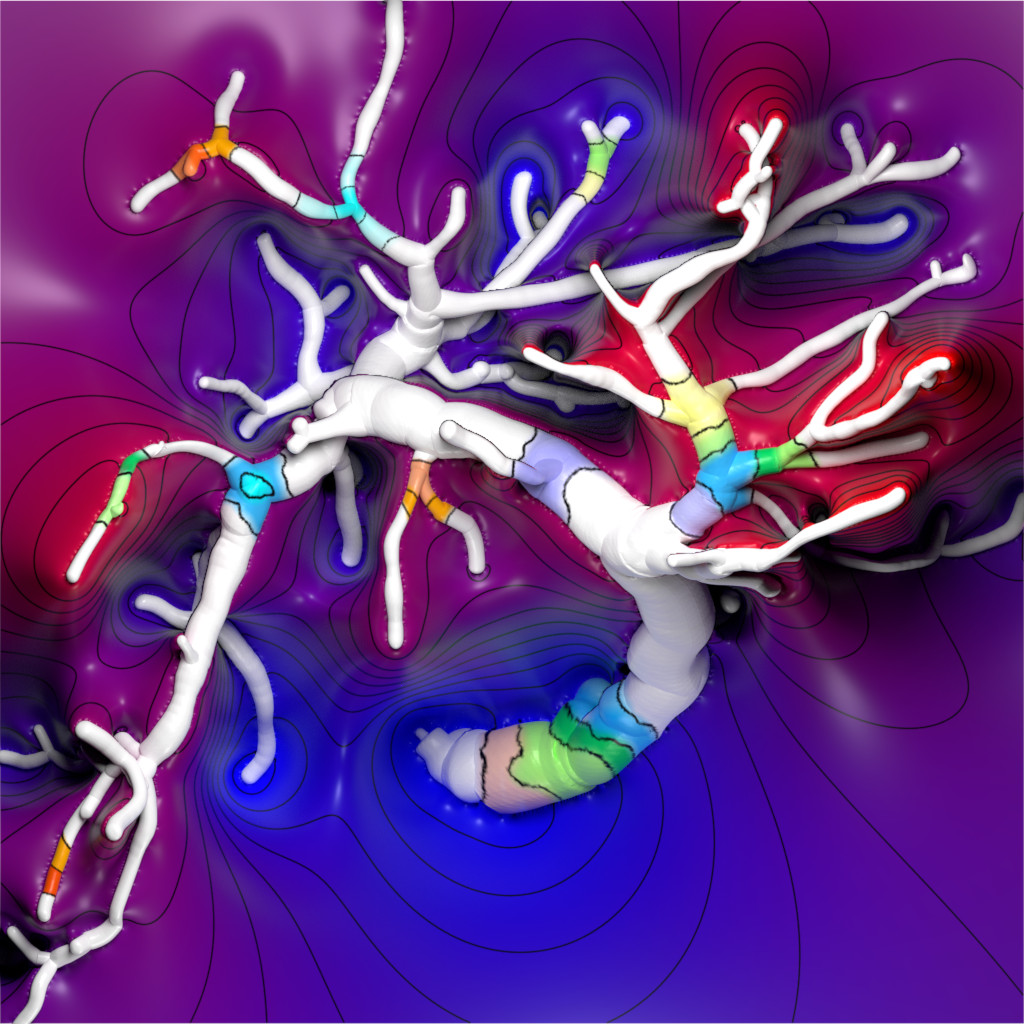}} \hfill
  \frame{\includegraphics[width=0.49\linewidth]{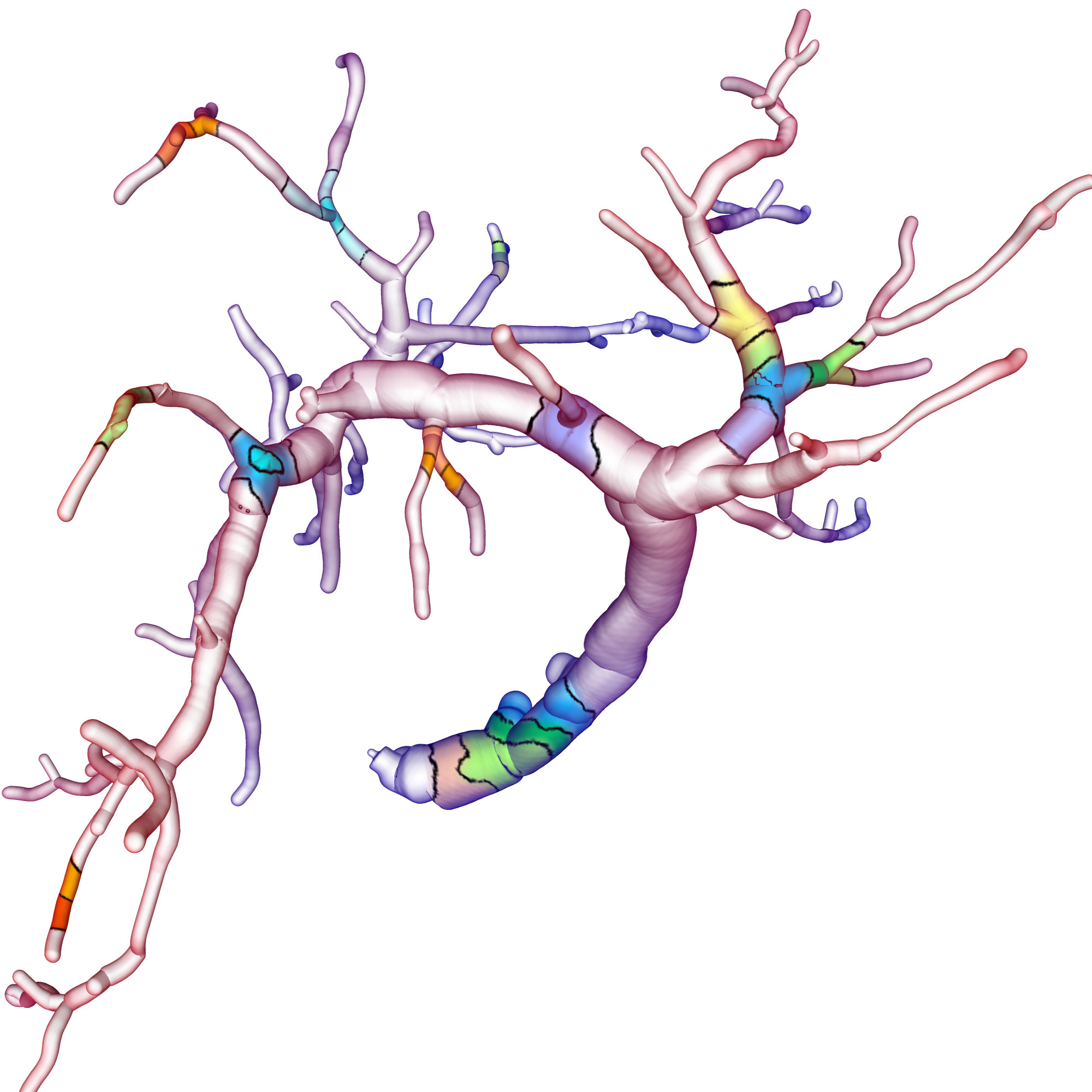}}
  \caption{
	Comparison of the same dataset rendered with the technique proposed by Behrendt et al.~\cite{Behrendt2017} (right, courtesy of Benjamin Behrendt) and ours (left) when showing additional parameters on the vessel's wall. With our method, depth and the relative position of the individual branches are effectively communicated, freeing the surface of the vessel to encode other relevant information.	In the state of the art, the pseudo-chromadepth colors have to be desaturated in order to not collide with the additional information presented in the same area.
	}
  \label{fig:comparison_behrendt}
\end{figure}

Figure~\ref{fig:comparison_behrendt} presents a comparison of the same model with similar camera parameters.
Although both, depth and additional parameters, are encoded by the two techniques in a single image, their method has to make some compromises since it uses the limited space of a vessel branch to encode both. The pseudo-chromadepth colors are not fully saturated in order to do not drive away the attention of the user from the additional parameter. Our method, on the other hand, does not have this drawback. Furthermore, we are able to convey depth through larger areas in the empty space between the vessel, which does not interfere with the encoding of additional parameters on the surface of the vessels and does not require to reduce the saturation of the colors. As these areas were previously unused, they do not compromise the actual visualization.

\subsection{Limitations}\label{subsec:limitations}
Our conducted evaluation indicates that the perceptual power of VSS is on par with the state-of-the-art, while it does not interfere with the color information of the vessels. Nevertheless, during the design of VSS we have also identified limitations which we would like to discuss in this section.

\noindent\textbf{Necessary void space.} The first and most obvious drawback is the necessity of void space between geometries of interest. Vascular structures satisfy this requirement quite well and are therefore our main application scenario. However, other medical assessment procedures are also suited for our visualization technique.
Rib cage data sets, the analysis of needle placements, or targeting tasks for radiofrequency tumor ablation where a probe has to be placed inside the target tumor without damaging surrounding structures can also benefit from enhanced depth and shape perception. While we expect that VSS work for such structures rather well, we do not expect such an improved performance when applied to more dense data sets.

\noindent\textbf{Background structures.}
The second scenario in which our method does not work is the case if an object shares no contour with the background and is therefore not attached to the VSS.
This happens if a structure lies completely in front of another one and gets occluded from the background side.
In this case, only occlusion and shading cues support the user.

\section{Conclusions and Future Work}\label{sec:conclusions}
Within this paper we have presented VSS, a novel technique to visualize complex vascular structure. VSS have been developed with our proposed vessel visualization design goals in mind, such that they communicate shape and depth while still allowing for the communication of functional parameters without requiring instrumentation of the user. They further work on static images, such as frequently used in medical reports, and do not require explanations to be interpretable. After introducing the computation of VSS, we have discussed the importance of depth anchoring, which enables the association of VSS and vessel trees. We have further shown how well-known depth enhancement techniques can be combined with VSS, whereby VSS serve as a canvas for these techniques. We could further show that the technical realization of VSS is simple, and that they can even be realized on the GPU in order to achieve interactive frame rates. The results of a user study conducted with 20 participants indicate that VSS help to improve the depth perception of complex vascular structure, by still supporting the visualization of functional parameters overlaid over the vessel's geometry. To our knowledge VSS is the first vessel visualization technique, that has these capabilities.

In the future, we plan to further investigate which set of parameters works best for VSS to achieve a maximum performance in depth enhancement. Eye tracking can therefor be used to analyze if VSS can function as passive cue or have to be directly perceived. Furthermore, we are considering to utilize VSS to communicate further information, such as functional parameters or external text labels. When combining such information with are pure iso-line depth enhancement, we expect that it can be beneficial. Furthermore, we are interested in investigating how much void space is actually needed to be useful for the user. When having such knowledge, it would become possible to predict for which other geometric structures VSS might be beneficial.

\acknowledgments{
This work was partially funded by the Deutsche Forschungsgemeinschaft~(DFG) under grant RO~3408/3-1~(Inviwo).
The authors thank the Ulm University Center for Translational Imaging MoMAN for its support.
}

\bibliographystyle{abbrv-doi}

\bibliography{vis18voidspacesurfaces}
\end{document}